\documentclass[a4paper,fleqn,usenatbib]{mnras}

\usepackage[T1]{fontenc}
\usepackage{ae,aecompl}

\usepackage{graphicx}
\usepackage{txfonts}
\usepackage{natbib}
\usepackage{longtable}
\usepackage{amssymb}
\usepackage{comment}
\newcommand{\kms}{km\,s$^{-1}$}

\title[The magnetic triple stellar system HD\,164492C.]{B fields in OB stars (BOB): The magnetic triple stellar system HD\,164492C in the Trifid nebula.\thanks{Based on observations obtained in the framework of the ESO Prgs.~191.D-0255(B,F,H,I,J), 093.D-0267(B), 091.C-0713(A), 088.D-0064(A), and 070.D-0191(A).}}

\author[J. F. Gonz\'alez et al.]
{J. F. Gonz\'alez$^{1}$\thanks{E-mail:{\tt fgonzalez@icate-conicet.gob.ar}}, 
S. Hubrig$^{2}$,  
N. Przybilla$^{3}$, 
T. Carroll$^{2}$,  
M.-F. Nieva$^{3}$,  
I. Ilyin$^{2}$, \and
S. J\"arvinen$^{2}$,
T. Morel$^{4}$, 
M. Sch\"oller$^{5}$, 
N. Castro$^{6,7}$,
R. Barb\'a$^{8}$,
A. de Koter$^{9,10}$, \and
F. R. N. Schneider$^{11}$,
A. Kholtygin$^{12}$,
K. Butler$^{13}$,
M. E. Veramendi$^{14}$,
N. Langer$^{7}$, \and
and the BOB collaboration
\\
$^{1}$Instituto de Ciencias Astron\'omicas, de la Tierra y del Espacio, Casilla de Correo 49, 5400 San Juan, Argentina\\
$^{2}$Leibniz-Institut f\"ur Astrophysik, An der Sternwarte 16, 14482 Potsdam, Germany\\
$^{3}$Institut f\"ur Astro- und Teilchenphysik, Universit\"at Innsbruck, Technikerstr. 25/8, 6020 Innsbruck, Austria\\
$^{4}$Space sciences, Technologies and Astrophysics Research (STAR) Institute, Universit\'e de Li\`ege, Quartier Agora, All\'ee du 6 Ao\^ut 19c, B\^at. B5C, B4000-Li\`ege, Belgium \\
$^{5}$European Southern Observatory, Karl-Schwarzschild-Str.~2, 85748 Garching, Germany\\
$^{6}$Department of Astronomy, University of Michigan, 1085 S. University Avenue, Ann Arbor, MI 48109-1107, USA\\
$^{7}$Argelander-Institut f\"ur Astronomie, Universit\"at Bonn, Auf dem H\"ugel 71, 53121 Bonn, Germany\\
$^{8}$Departamento de F\'isica y Astronom\'ia, Universidad de La Serena, Av. Cisternas 1200 Norte, La Serena, Chile\\
$^{9}$Anton Pannekoek Institute for Astronomy, University of Amsterdam, Science Park 904, PO Box 94249, 1090 GE, Amsterdam, The Netherlands\\
$^{10}$Instituut voor Sterrenkunde, KU Leuven, Celestijnenlaan 200D, 3001, Leuven, Belgium\\
$^{11}$Department of Physics, Denys Wilkinson Building, Keble Road, Oxford, OX1 3RH, United Kingdom\\
$^{12}$Chair of Astronomy, Astronomical Institute, St. Petersburg State University, Universitetski pr. 28, 198504, St. Petersburg, Russia\\
$^{13}$Universit\"ats-Sternwarte M\"unchen, Scheinerstr. 1, 81679 M\"unchen, Germany\\
$^{14}$Complejo Astron\'omico El Leoncito, Casilla de correo 467, 5400 San Juan, Argentina}
\date{Accepted XXX. Received YYY; in original form ZZZ}

\pubyear{2016}

\begin{document}
\label{firstpage}
\pagerange{\pageref{firstpage}--\pageref{lastpage}}
\maketitle

\begin{abstract}
HD\,164492C is a spectroscopic triple stellar system that has been recently detected to possess 
a strong magnetic field. 
We have obtained high-resolution spectroscopic and spectropolarimetric observations over  a
timespan of two years and derived  physical, chemical, and magnetic properties for this object. 
The system is formed by an eccentric close spectroscopic binary (Ca1-Ca2) with a period of 12.5\,days,
 and a massive tertiary Cb. 
We calculated the orbital parameters of the close pair, reconstructed the spectra of the three components, and determined atmospheric parameters and chemical abundances by spectral synthesis.
From spectropolarimetric observations, multi-epoch  measurements of the longitudinal magnetic fields were obtained.
The magnetic field is strongly variable on timescales of a few days, with a most probable period
in the range of 1.4--1.6\,days. 
Star Cb with $T_\mathrm{eff}$\,$\sim$\,25\,000\,K is the apparently
fastest rotator and the most massive star of this triple system and has
anomalous chemical abundances with a marked overabundance of helium,
0.35$\pm$0.04 by number.
We identified this star as being responsible for the observed magnetic field, although 
the presence of magnetic fields in the components of the Ca pair cannot be excluded.
Star Ca1 with a temperature of about 24\,000\,K
 presents a normal chemical pattern, while the least massive star Ca2 is a mid-B type star 
 ($T_\mathrm{eff}$\,$\sim$\,15\,000\,K) with an overabundance of silicon.
 The obtained stellar parameters of the system components suggest a distance of 1.5\,kpc and an age of  10--15\,Myr.
\end{abstract}


\section{Introduction}

The multiple star HD\,164492 (=ADS 10991) is a trapezium-like stellar system
belonging  to the Trifid Nebula, an active star forming region.
The nebula is ionised by the O7.5~Vz \citep{2014ApJS..211...10S} star HD\,164492A, which is the central object of this multiple. 
The central part of the system is formed by five bright ($V=7.4-12.2$\,mag) visual components in a non-hierarchical configuration \citep[A to E;][]{1999A&AS..134..129K}.
The relevance of studying this multiple has increased following the recent discovery
of a magnetic field of about 500\,G in one of its components, the subsystem HD\,164492C \citep{2014A&A...564L..10H}, by the BOB (``B fields in OB stars'') collaboration \citep{2014Msngr.157...27M,2015IAUS..307..342M}.
In addition, \citet{2014A&A...564L..10H} reported that HD\,164492C is an early-B type double-lined spectroscopic binary, and possibly a triple. 
No quantitative analyses of the components of HD\,164492C is
available in the literature so far.

The main goal of the present work is to describe the physical and chemical properties of this interesting system
as completely as  possible, in particular its magnetic field,
in order to contribute to the understanding of the conditions under which 
magnetic fields are developed in massive stars.
Only a small fraction of stars (5--7\%) with
radiative envelopes possess strong large-scale  organised magnetic
fields \citep{2013EAS....64...67G,2015IAUS..307..342M}, 
which can be generated during the star formation process
\citep[fossil fields, e.g.][]{1982ARA&A..20..191B,2001ASPC..248..305M}, 
or dynamo action taking place in the rotating stellar cores, or they could be products of a merger process. 
While the first two scenarios are unable to explain a number of observational 
phenomena \citep[e.g.][]{2015SSRv..191...77F}, the magnetic fields could form when
two proto-stellar objects merge late on their approach to the main
sequence and when at least one of them has already acquired a
radiative envelope \citep{2009MNRAS.400L..71F}. 
The origin of magnetic fields in massive stars is still a major unresolved problem in astrophysics. 

Sections \ref{sec:obs} to \ref{sec:orb}  describe the observations, the reconstruction of
the spectra of the stellar components, and the orbital analysis of the close binary subsystem. 
The  physical parameters and chemical abundances are derived in Sect. \ref{sec:abun} and \ref{sec:param}, while the magnetic field 
is  discussed in Sect. \ref{sec:magfield}.
The main results are summarised in Sect. \ref{sec:disc}.
 
\section[]{Spectroscopic and spectropolarimetric observations}\label{sec:obs}

We have obtained high-resolution spectroscopic and spectropolarimetric observations over a timespan of
two years with the ESO spectrographs UVES, FEROS, and HARPS. 
The FEROS echelle spectrograph on the 2.2-m telescope at La~Silla produces spectra with
a resolving power of 48\,000 and a wavelength coverage from 3500 to 9200\,\AA{}.    
Our UVES observations cover two spectral regions: the blue spectrum between 
3300 and 4530\,\AA{} with a resolving power of $R=80\,000$ and the red spectrum covering 
the range from 5700 to 9450\,\AA{} with $R=110\,000$.
The polarimetric spectra of HARPS have a resolving power of $R=115\,000$ and cover
the spectral range from 3780 to 6910\,\AA{}, with a gap between 5259
and 5337\,\AA{}.
The four FEROS spectra were taken in August 2013, while the 15 UVES spectra
were taken between  April and August 2014.
The typical signal-to-noise ratio at $\lambda=4200$\,\AA{} is 280 for
the UVES spectra and 110 for the FEROS spectra.
With HARPS we acquired 11 spectropolarimetric observations in 5 observing runs in
June 2013, April 2014, March 2015, June 2015, and October 2015. 
 The signal-to-noise ratio in the
Stokes~$I$ spectra is about 300 per pixel above 4600\,\AA{},
decreasing in the blue to about 150 at 4000\,\AA{}. 

The basic data reduction and calibration was performed using specific ESO reduction pipelines.
UVES spectra were re-reduced using standard tasks of the NOAO/IRAF package.  

 In the red spectral region, the spectra required some processing to remove features 
not belonging to the stars: telluric and interstellar absorption 
lines, and, in the case of the HARPS spectra, nebular emission lines.
Emission features appear with different intensity in spectra
taken with the  fibre-fed spectrograph HARPS, while they are absent in observations
taken with the slit spectrograph UVES. The fact that these emissions are removed
through background subtraction  during the reduction process of UVES data, indicates that 
the origin of these emissions is not circumstellar but nebular, since any circumstellar
structure would have been included in the object extraction aperture.
We obtained an emission line master spectrum by averaging background-aperture spectra of several 
UVES observations. This emission line spectrum was scaled to mimic the intensity of emission
lines in each HARPS spectrum and then subtracted.

During this process we identified and measured all relevant nebular lines.
The line identification was done using the lists by \citet{2009elu..book....1S}.
The average radial velocity of nebular lines is $-4.8\pm1.8$\,\kms.
Most emission lines do not overlap crucial stellar lines, but 
the detection and removal of small He\,\textsc{i} emissions was particularly important to avoid 
the distortion of stellar He\,\textsc{i} line profiles.

 \section[]{Separation of spectral components}\label{sec:sep}

Line profiles of HD\,164492C show clear variations on time-scales of a few days.
The majority of the line profiles presents a complex structure suggesting the existence of more than
two spectroscopic components. 
We propose as the most simple interpretation, the presence of a fixed broad-line
spectrum superimposed with a double-lined spectroscopic binary. 
The results of the spectral disentangling performed under this hypothesis and the
modelling of composite spectra with spectral synthesis confirmed that this system is indeed a hierarchical triple.
The presence of a third source is also supported by the fact that HD\,164492C is a visual 
binary with a separation
of 0.08\arcsec \citep{2016BAAA...58..105G,2005AJ....130.1171Y}. 
This visual pair can be confidently considered as physically bound, since the
probability of chance alignment for two stars of magnitude $V$=9.0--9.5 mag 
($J$=8.6--9.2 mag) is very low.  
From the two-dimensional star density, derived using the 2MASS Catalogue, we estimated  
that the probability of having a companion brighter than $J$=10 mag at an
angular distance smaller than 0.08\arcsec is about 1:$1.3\times10^ 6$. 
Following the conventional nomenclature of multiple stars, we named the companions of the
close binary as HD\,164492Ca1 and HD\,164492Ca2, and the more distant star with the broad-line spectrum as HD\,164492Cb.  

We separated the three spectral components applying the method of \citet{2006A&A...448..283G}, and
assuming that the third star (Cb) has constant radial velocity.
Considering the different wavelength coverage of the HARPS and UVES spectra, we performed the 
calculations separately in three spectral regions: 
3700--4530\,\AA{} (UVES+HARPS), 3800--5250\,\AA{} (HARPS), and 5680--6800\,\AA{} (UVES+HARPS).  
From the composite spectra we made a first estimation of the spectral types (B1 for components Ca1 and Cb, 
and B4-B6 for Ca2) and according to this we selected 
spectra of reference stars as a starting point for the iterative disentangling process.
We used high-resolution archival spectra of slowly-rotating 
stars HR\,2222 (B1\,V) for Ca1 and Cb and HR\,1288 (B4\,V) for Ca2. These spectra have been taken with the 
UVES and FEROS spectrographs as part of other  observing programmes. 
The reference spectra were scaled and convolved with theoretical rotational profiles to mimic the lines of the components of HD~164492C.

\begin{figure*}
   \centering
 \begin{minipage}[]{8.5cm}
  \includegraphics[bb= 22 32 570 735,height = 8.5cm, width=9cm,angle=0]{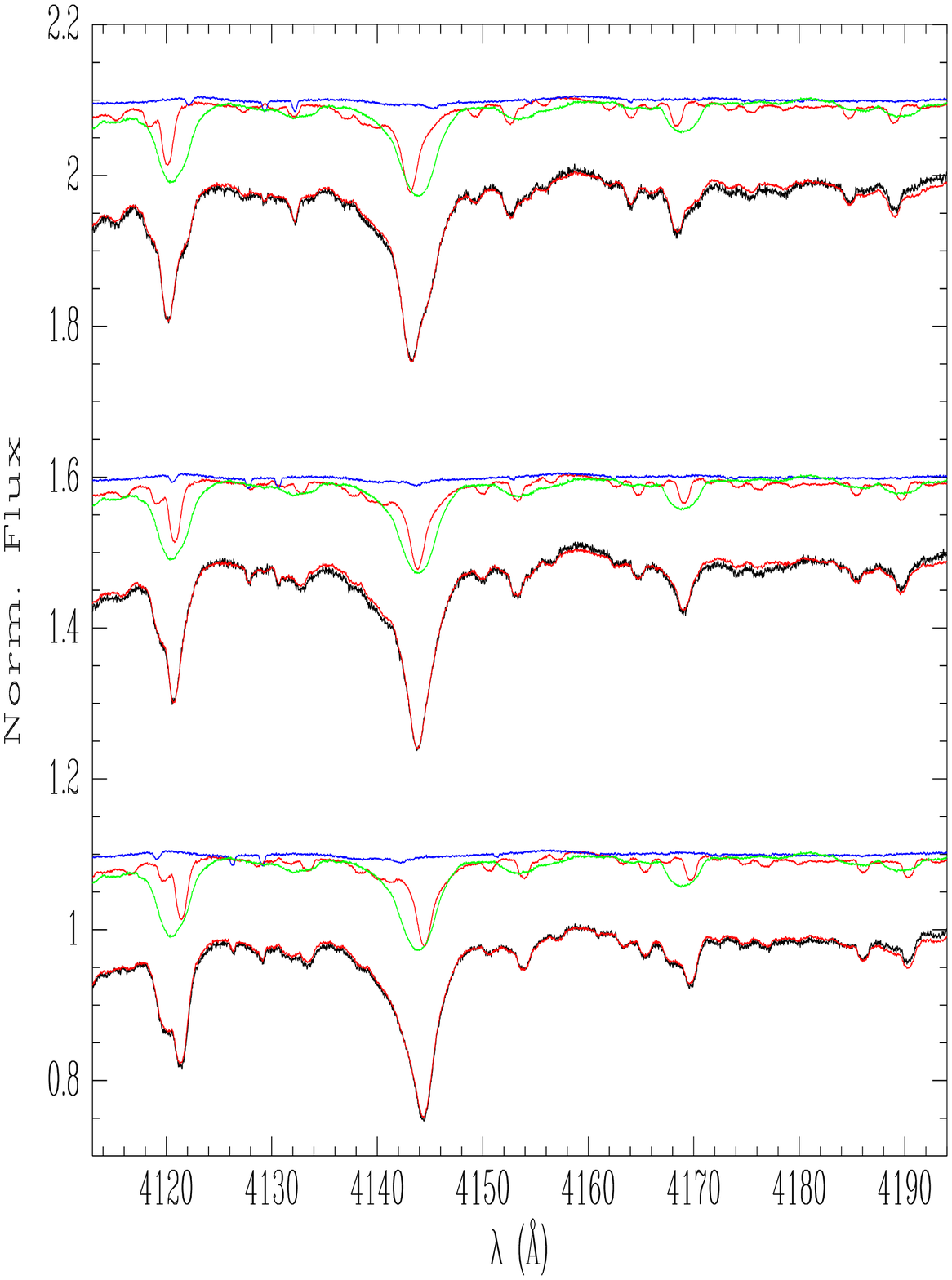} 
\end{minipage}
\begin{minipage}[]{8.5cm}
   \includegraphics[bb= 2 32 550 735,height = 8.5cm, width=9cm,angle=0]{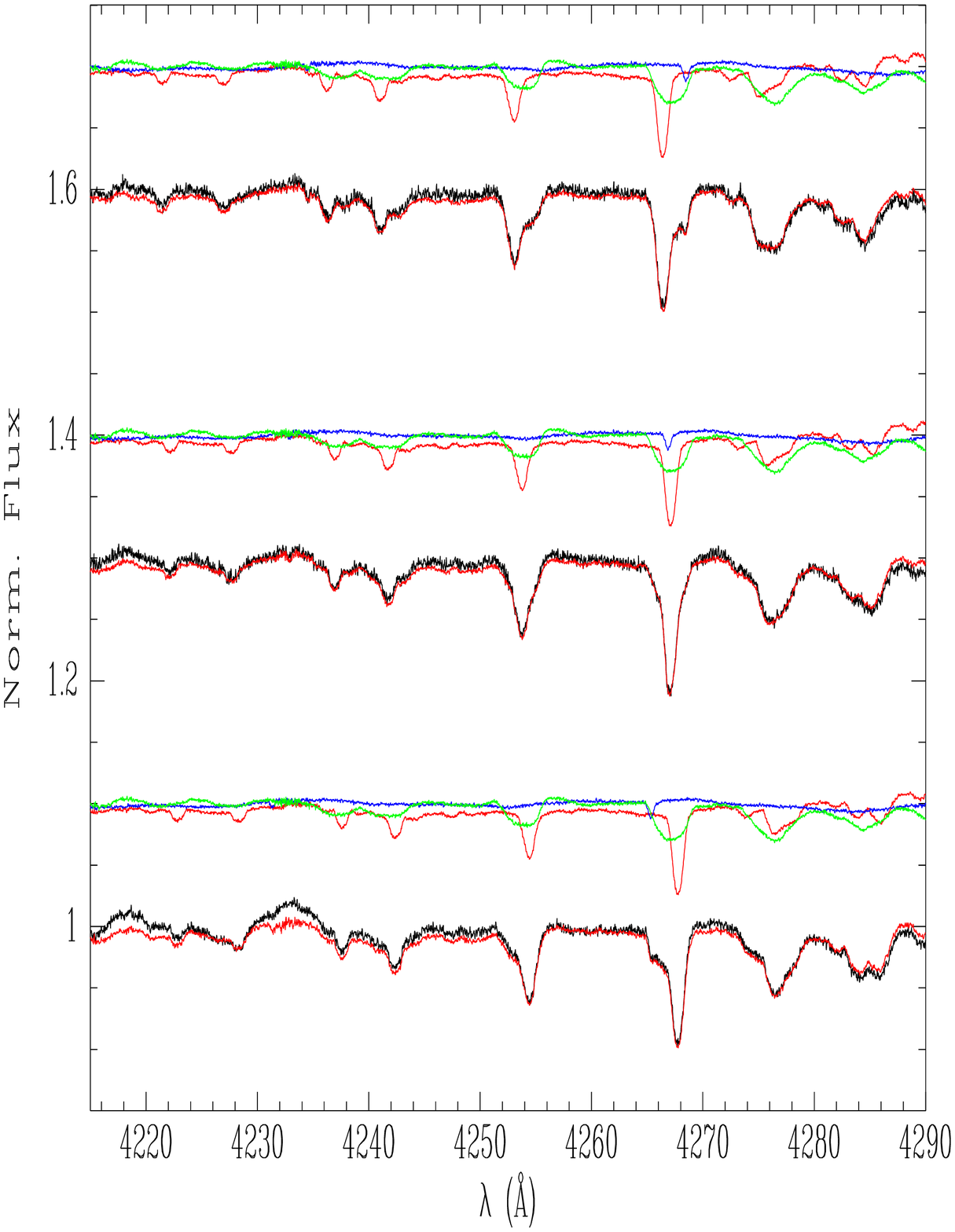} 
\end{minipage} 
    \includegraphics[bb= 2 32 550 735,height = 8.5cm, width=9cm,angle=0]{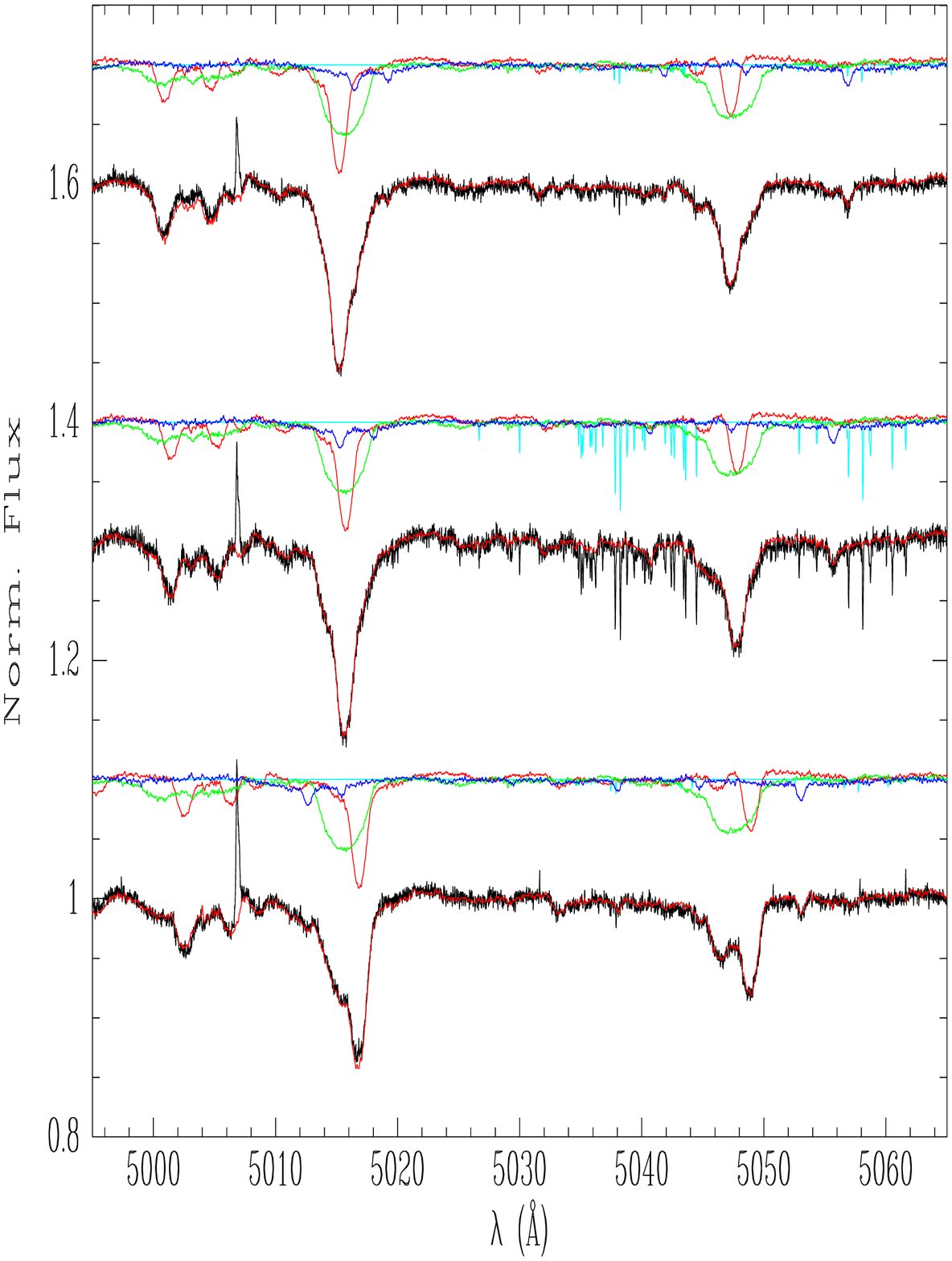}%
   \caption{Selected regions in the UVES spectra showing the separation of the spectral components at three different phases. The two upper panels correspond to UVES spectra with orbital phases 0.14, 0.35, and 0.81
   (from bottom to top) and the lower panel to HARPS spectra with phases 0.02, 0.33, and 0.56.
   For each phase we show in the upper part the reconstructed spectra of the components Cb (green), Ca1 (red), and Ca2 (blue), and
   below the observed spectrum (small black dots) overplotted with the sum of the three reconstructed spectra (red line). 
In the lower panel modelled telluric lines are shown in light blue.
The nebular $[$O\,\textsc{iii}$]$ emission line is visible at 5007\,\AA{}.}\label{fig:separa}
   \end{figure*}

To recover the intrinsic intensities of the spectra of the individual components, 
it is necessary to scale the separated spectra according to the flux contribution of each star. 
To this aim, we derived from the spectral types a first estimate of $T_\mathrm{eff}$ 
and $\log g$. Then we estimated 
the flux contribution of each component as a function of wavelength using synthetic spectra 
from the grid calculated by \citet{2007ApJS..169...83L} and the stellar radii estimated
from the Geneva stellar models \citep{2012A&A...537A.146E} assuming that all three stars have 
the same age and requiring the masses of stars Ca1 and Ca2 to be consistent with the spectroscopic mass ratio.
In a later stage, the relative flux of the components was recalculated using synthetic spectra
with the appropriate atmospheric parameters and chemical abundances (see Sect.~\ref{sec:abun}).

The noise level of the separated spectra is about 1/1300 of the original continuum at $\lambda$ 4200\,\AA{} 
in the first spectral region, about 1/650 at 5000\,\AA{},  and 1/800 in the red region. 
The final signal-to-noise  of the reconstructed spectra depends also on 
the relative flux contribution of each star, which are approximately 0.55--0.60 for Cb, 0.35--0.40 for Ca1, 
and 0.05--0.07 for Ca2 (see Sect.~\ref{sec:abun}).
Thus, after scaling, the  separated spectra have a signal-to-noise ratio of the order of 500 for the most
massive stars and of only 50 for the star Ca2. 

\begin{table*}
 \centering
 \begin{minipage}{0.7\textwidth}
  \caption{Radial velocities of the spectroscopic binary HD\,164492Ca. Column 1: Heliocentric Julian date; Column 2: orbital phase calculated with respect to the periastron passage $T_\pi$ (see Table~\ref{tab:orb}); Column 3 and 5: radial velocities for stars Ca1 and Ca2; Columns 4 and 6:  residuals observed-minus-calculated; Column 7: Spectrograph.}\label{tab:rvs}
\centering
  \begin{tabular}{@{}ccr@{ $\pm$ }rrr@{ $\pm$ }rrc}
  \hline
    HJD  & $\phi$ & \multicolumn{2}{c}{RV$_\rmn{Ca1}$}& \multicolumn{1}{c}{(O-C)$_1$} &
                            \multicolumn{2}{c}{RV$_\rmn{Ca2}$} & \multicolumn{1}{c}{(O-C)$_2$} & \multicolumn{1}{c}{Instrument}\\ 
         	    &		    &	\multicolumn{2}{c}{\kms} & \multicolumn{1}{c}{\kms}  & \multicolumn{2}{c}{\kms}  & \multicolumn{1}{c}{\kms}  & \\ \hline
2\,456\,445.7946&  0.5559& $-$28.4&   1.0&    1.7&     44.1&   0.6&    0.0&   HARPS\\
2\,456\,446.7934&  0.6355& $-$36.6&   0.9&    2.0&     62.3&   0.7& $-$0.9&   HARPS\\
2\,456\,522.6844&  0.6898& $-$42.7&   1.4&    1.2&     75.0&   1.9&    0.0&   FEROS\\
2\,456\,523.7039&  0.7711& $-$49.9&   1.2&    0.4&     88.9&   1.9& $-$0.7&   FEROS\\
2\,456\,524.6224&  0.8444& $-$52.9&   1.6& $-$0.5&     98.8&   2.6&    4.5&   FEROS\\
2\,456\,525.6624&  0.9273& $-$32.3&   1.4&    2.6&     59.1&   2.0&    4.3&   FEROS\\
2\,456\,770.7700&  0.4808& $-$22.5&   1.0& $-$1.2&     24.0&   0.9& $-$0.2&   HARPS\\
2\,456\,771.7850&  0.5618& $-$28.3&   0.9&    2.5&     46.3&   0.9&    0.7&   HARPS\\
2\,456\,772.7707&  0.6404& $-$39.4&   0.8& $-$0.2&     63.1&   0.8& $-$1.2&   UVES\\
2\,456\,777.8085&  0.0423&    73.9&   0.9& $-$1.2& $-$192.7&   0.7&    0.2&   UVES\\
2\,456\,779.8951&  0.2088&    25.4&   1.0& $-$0.3&  $-$83.5&   0.9& $-$2.0&   UVES\\
2\,456\,786.8817&  0.7661& $-$51.2&   0.8& $-$1.2&     88.2&   0.5& $-$0.6&   UVES\\
2\,456\,837.6044&  0.8125& $-$53.4&   0.8& $-$1.2&     94.3&   0.5&    0.3&   UVES\\
2\,456\,869.5077&  0.3576&  $-$4.2&   0.9& $-$0.1&  $-$12.7&   0.5&    1.9&   UVES\\
2\,456\,879.6971&  0.1705&    36.1&   0.9& $-$0.2& $-$105.2&   0.7&    0.4&   UVES\\
2\,456\,883.4874&  0.4729& $-$22.5&   0.9& $-$2.2&     20.7&   0.4& $-$1.2&   UVES\\
2\,456\,887.4936&  0.7925& $-$52.4&   0.9& $-$0.9&     93.3&   0.5&    1.1&   UVES\\
2\,456\,894.6261&  0.3614&  $-$3.8&   0.9&    0.9&  $-$12.3&   0.6&    0.9&   UVES\\
2\,456\,898.5681&  0.6759& $-$43.8&   0.9& $-$1.2&     71.8&   0.7& $-$0.3&   UVES\\
2\,456\,919.5607&  0.3506&  $-$3.1&   1.0& $-$0.2&  $-$17.0&   0.5&    0.1&   UVES\\
2\,456\,923.5111&  0.6658& $-$41.7&   0.9& $-$0.1&     67.9&   1.0& $-$2.0&   UVES\\
2\,456\,926.5385&  0.9073& $-$42.8&   0.8&    0.7&     72.2&   0.7& $-$2.0&   UVES\\
2\,456\,929.4996&  0.1435&    43.8&   0.8& $-$1.2& $-$126.4&   0.9& $-$1.4&   UVES\\
2\,457\,091.8426&  0.0944&    59.1&   1.3& $-$3.6& $-$163.8&   1.0&    1.3&   HARPS\\
2\,457\,092.8282&  0.1730&    35.9&   0.9&    0.4& $-$103.4&   0.6&    0.5&   HARPS\\
2\,457\,093.8478&  0.2544&    17.2&   0.7&    2.3&  $-$57.5&   0.6& $-$0.1&   HARPS\\
2\,457\,094.8153&  0.3316&     3.2&   0.7&    3.0&  $-$24.0&   0.7&    0.3&   HARPS\\
2\,457\,178.6948&  0.0231&    69.5&   1.4&    0.6& $-$180.9&   1.1& $-$2.0&   HARPS\\
2\,457\,179.7426&  0.1067&    60.2&   1.4&    2.0& $-$156.5&   0.9& $-$1.8&   HARPS\\
2\,457\,317.5185&  0.0977&    64.6&   2.0&    3.1& $-$163.9&   2.0& $-$1.7&   HARPS\\
\hline
\end{tabular}
\end{minipage}
\end{table*}

As illustration, Figure~\ref{fig:separa} shows for three different orbital phases how the reconstructed 
spectra combine to reproduce the observed profiles.
For the star Ca2, lines of Si\,\textsc{ii} ($\lambda\lambda$ 4128, 4130, 5041, 5056), Fe\,\textsc{ii} ($\lambda\lambda$ 4233, 5018),
C\,\textsc{ii} ($\lambda$ 4267), and He\,\textsc{i}  ($\lambda\lambda$ 4121, 4144, 5015, 5047) can be distinguished. For the most massive stars, besides the mentioned lines of C\,\textsc{ii} and He\,\textsc{i}, the most noticeable lines in the plot correspond to
Al\,\textsc{iii} ($\lambda$ 4150), Fe\,\textsc{iii} ($\lambda$ 4165), and O\,\textsc{ii} ($\lambda\lambda$ 4133, 4153, 4169, 4185, 4190).
A complete spectral atlas of the component spectra with the line identification
 is included in the Appendix (available online).
We conclude that the three mentioned spectral components represent satisfactorily
the observed spectral content and variations.

The reconstructed spectrum of Ca1, the primary companion of the close binary, shows
a morphology typical of a B1\,V star, with lines of He\,\textsc{i}, C\,\textsc{ii}, N\,\textsc{ii}, 
O\,\textsc{ii}, Al\,\textsc{iii}, Si\,\textsc{iii-iv}, S\,\textsc{ii-iii}, and Fe\,\textsc{iii}.
The spectrum of Cb seems to be very similar to Ca1, but with broader
metal lines because of a higher rotational velocity. Remarkable
are the unusually strong He\,\textsc{i} lines, indicating a He-rich
nature for this star.
The faint star Ca2 is a mid-B type star: only a few strong lines of  He\,\textsc{i}, 
C\,\textsc{ii},  Mg\,\textsc{ii}, Si\,\textsc{ii}, S\,\textsc{ii}
and Fe\,\textsc{ii} can be identified in its spectrum (see Appendix).

 \section[]{Orbital elements}\label{sec:orb}

Radial velocities of the close binary Ca were measured by cross-correlations
during the spectral separation process. 
FEROS spectra, which were not considered in the reconstruction of component spectra  because of
their lower resolution, were included in the radial velocity measurements. 
We used as templates the observed spectra of reference stars mentioned in the previous
section: HR\,2222 for Ca1 and HR\,1288 for Ca2. 
 Before carrying out the cross-correlations, radial velocities of template spectra 
were corrected by measuring several tens of metallic spectral lines.
Table~\ref{tab:rvs} lists the measured radial velocities.
Orbital phase zero corresponds to the periastron passage.

During the spectral disentangling only one fixed value for the radial velocity
of star Cb was fitted. Afterwards, to look for possible variations in the spectrum
of this star, we removed the spectra of Ca1 and Ca2 from the observed spectra
using the reconstructed spectra of these components, and measured the
resulting spectra, which would be individual spectra of star Cb.
The velocity of this star is constant within uncertainties, with a mean
value of $-6.3$\,\kms\ and a standard deviation of 2.8\,\kms, which is
lower than the typical measurement errors. 
The mean radial velocity of Cb agrees with the barycentric velocity 
of the binary Ca and that of the Trifid nebula ($-4.8\pm1.8$\,\kms, see Sect.~\ref{sec:obs}).

\begin{figure}
   \centering
 \includegraphics[bb= 2 20 550 737, width=7cm,height=1.03\linewidth,angle=-90]{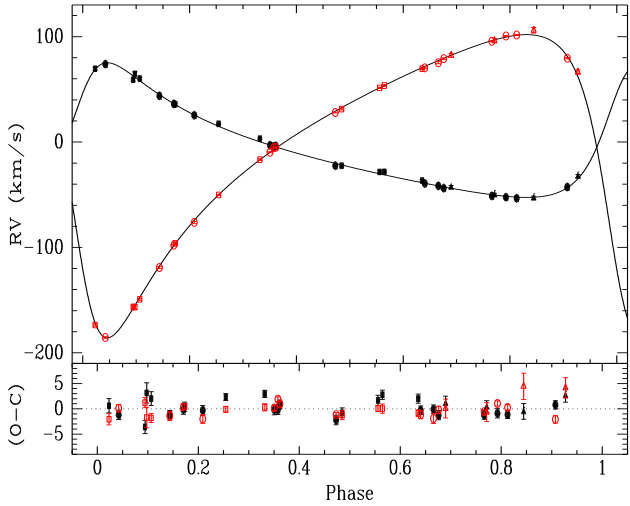}
   \caption{Radial velocity curves of HD\,164492Ca. 
      The bottom panel shows residuals (observed-minus-calculated).
   Filled (open) symbols correspond to the primary (secondary) star,
   circles indicate measurements with UVES, squares with HARPS, and
   triangles with FEROS.}
         \label{fig:rvcurve}
   \end{figure}

\begin{table}
\caption{Orbital parameters of the binary HD\,164492Ca. }
\label{tab:orb}
\centering
\begin{tabular}{lcr@{ $\pm$ }l} 
 \hline
Parameter	&	Units	&	 \multicolumn{2}{c}{Value }	\\ \hline
$P$			&	d			&	12.5349	&	0.0008		\\
$T_\mathrm{I}$	&	d		&	2\,456\,692.94	&	 0.03		\\
$T_\pi$		&	d			&	2\,456\,689.54	&	0.03		\\
$K_\rmn{1}$ &	\kms &	63.8	&	0.8		\\
$K_\rmn{2}$	&	\kms &	143.8	&	0.7	\\
$\omega$	&		rad		&	5.29	&	0.01\\
$e$			&				&	0.532	&	0.004	\\
$q$			&       		&	0.444	&	0.006	\\
$a_\rmn{1}\sin i$&$R_\odot$	&	13.4	&	0.2		\\
$a_\rmn{2}\sin i$&$R_\odot$	&	30.2	&	0.2	\\
$a\sin i$		&$R_\odot$	&	43.5	&	0.2	\\
$M_\rmn{1}\sin^3 i$&$M_\odot$&	4.89	&	0.06	\\
$M_\rmn{2}\sin^3 i$	&$M_\odot$&	2.17	&	0.04	\\
$\gamma$	&\kms	&     $-$7.3	&	0.3		\\
$n$		&			&	 \multicolumn{2}{c}{30}		\\
$\sigma_\rmn{1}$&\kms &	 \multicolumn{2}{c}{1.65}\\
$\sigma_\rmn{2}$	&\kms 	&	 \multicolumn{2}{c}{1.61}\\
\hline
\end{tabular} 
\end{table}

A Keplerian orbit was fitted to the observed radial velocities  by least squares.
Figure~\ref{fig:rvcurve} shows the radial velocity curves and Table~\ref{tab:orb} lists
the fitted orbital parameters: period $P$, time of periastron passage $T_\pi$, 
radial velocity amplitudes $K_\mathrm{1}$ and $K_\mathrm{2}$,  centre-of-mass velocity $\gamma$,
argument of periastron $\omega$, and eccentricity $e$. We list also the following 
derived parameters: time of primary conjunction $T_\mathrm{I}$, 
projected orbital semiaxis $a\sin i$, and minimum masses $M\sin^3 i$. 
The last three rows are the number of observations and the global RMS of the residuals
for the primary ($\sigma_\rmn{1}$) and the secondary ($\sigma_\rmn{2}$).
In fact, standard deviations differ for the three instruments, being
about 1.1\,\kms\ for UVES, 1.7\,\kms\ for HARPS, and 2.5\,\kms\ for FEROS. 
These values are somewhat larger than the formal measurement errors, and are probably 
a better estimate of the actual velocity uncertainties.

\section{Spectral analysis}\label{sec:abun}
The determination of physical and chemical properties of stellar atmospheres 
through spectral analysis and modelling is more complicated in the case of spectroscopic 
multiple systems,  because more parameters need to be accounted for. 
A comprehensive solution requires the physical
model of the system to match the observed composite and disentangled spectra and to provide
consistent ages, spectroscopic distances, the mass ratio for Ca2/Ca1 (from the
orbital solution) and flux scaling factors for
the components {\em simultaneously}. Observationally, this is
complicated by the fact that low-frequency modulations are not recovered 
by disentangling techniques based on relative Doppler shifts. 
For that reason broad features like hydrogen lines and broad He 
lines, whose profiles are significantly
wider than the shifts due to the orbital motion, cannot be fitted individually for each component.
 Moreover, relative fluxes are in principle unknown. Unless they are adopted 
from a different information source -- as e.g.~in detached eclipsing binaries --, 
they are additional parameters,  however mutually dependent on
stellar effective temperatures and radii. These facts make the spectral analysis of 
the various stellar components to be interdependent.

\begin{table}
\caption[]{Model atoms for non-LTE calculations. Updated models
as described by \citet{2012A&A...539A.143N} are marked with *. \\[-3mm] }\label{tab:atoms}
\setlength{\tabcolsep}{.35cm}
\begin{tabular}{ll}
\hline
\footnotesize
            Ion     &  Model atom \\ \hline \\[-3mm]
 H\,\textsc{i}      &  \citet{2004ApJ...609.1181P} \\
He\,\textsc{i/ii}   &  \citet{2005AA...443..293P} \\
 C\,\textsc{ii/iii} &  \citet{2006ApJ...639L..39N,2008AA...481..199N}\\
 N\,\textsc{ii}     &  \citet{2001AA...379..955P}*\\
 O\,\textsc{i/ii}   &  \citet{2000AA...359.1085P}, \citet{1988AA...201..232B}*\\
Ne\,\textsc{i}      &  \citet{2008AA...487..307M}*\\
Mg\,\textsc{ii}     &  \citet{2001AA...369.1009P}\\
Al\,\textsc{iii}    &   Przybilla (in prep.)\\
Si\,\textsc{ii-iv}  &   Przybilla \& Butler (in prep.)\\
 S\,\textsc{ii/iii} &  \citet{1996AA...311..661V}*\\
Fe\,\textsc{ii/iii} &  \citet{1998ASPC..131..137B}, \citet{2006AA...457..651M}*\\
            \hline\\[-3mm]
           \end{tabular}
\end{table}

 We used a hybrid non-LTE approach for the quantitative analysis
of the composite and separated spectra, in analogy to the methods
described by \citet{2007A&A...467..295N,2012A&A...539A.143N} and
\citet{2016A&A...587A...7P} for the
study of single (He-strong) B-type stars, extending them here where
necessary for the multiple star case. In brief,
this approach is based on plane-parallel, hydrostatic, chemically
homogeneous and line-blanketed model atmospheres that were computed with the code {\sc Atlas9} 
\citep{1993KurCD..13.....K} under the assumption of radiative and local thermodynamic
equilibrium (LTE). For the parameter range studied here, these are equivalent 
to hydrodynamic or hydrostatic line-blanketed non-LTE model
atmospheres within the line-formation regions
\citep{2011A&A...532A...2N,2007A&A...467..295N,2011JPhCS.328a2015P}. 
Non-LTE level populations and synthetic spectra were calculated with recent versions of the codes 
{\sc Detail} and {\sc Surface} \citep[both updated by one of us (KB)]{gid81,but_gid85}, 
employing comprehensive model atoms as summarised in Table~\ref{tab:atoms}. 

\begin{table*}
\centering
\caption{Atmospheric  and fundamental stellar parameters and chemical abundances 
(\,$\log X/\mathrm{H} + 12$) for the three components of HD\,164492C. 
Uncertainties are $\sim$0.2\,dex  for abundances.
Standard values (CAS) were taken from \citet{2012A&A...539A.143N}, except those marked
with asterisks, which are preliminary values from
\citet{2013EAS....63...13P}. Uncertainties for the CAS values are standard deviations.} 
\label{tab:atmpar}
\centering
\begin{tabular}{ccccr} \hline
\multicolumn{1}{c}{Parameter} &
\multicolumn{1}{c}{HD\,164492Cb} &
\multicolumn{1}{c}{HD\,164492Ca1} &
\multicolumn{1}{c}{HD\,164492Ca2} &
\multicolumn{1}{c}{CAS} \\
 \hline \smallskip
Spectral Type        & B1\,V/IV He-strong & B1\,V & B4-6\,V\\
$T_\mathrm{eff}$ (K) & $25\,000\pm500$& $24\,000\pm500$  & $15\,000\pm1000$  &\\
$\log g$ (cgs)       & $3.90\pm0.15$  & $4.00\pm$ 0.20  & 4.30$^a$ & \\
$\xi$ (km s$^{-1}$)  & $2\pm1$        & $2\pm1$        & $2\pm1$&\\
$v \sin i$ (km s$^{-1})$&$135\pm5$    & $48\pm5$       & $19\pm2$&\\
$\zeta$ (km s$^{-1}$)& {\ldots}       & $25\pm5$       & $12\pm2$&\\\hline%
$y$ (number fraction)& $0.35\pm0.04$  & $0.089\pm0.010$& $0.089\pm0.010$&$0.089\pm0.002$\\
C\,\textsc{ii/iii}   & 8.13           & 8.50           & {\ldots}&$8.33\pm0.04$\\
N\,\textsc{ii}       & 7.78           & 7.93           & {\ldots}&$7.79\pm0.04$\\
O\,\textsc{i/ii}     & 9.16           & 8.91           & {\ldots}&$8.76\pm0.05$\\
Ne\,\textsc{i}       & 7.59           & 8.15           & {\ldots}&$8.09\pm0.05$\\
Mg\,\textsc{ii}      & 7.36           & 7.62           & {\ldots}&$7.56\pm0.05$\\
Al\,\textsc{iii}     & 6.23           & 6.49           & {\ldots}&*$6.30\pm0.07$\\
Si\,\textsc{ii-iv}   & 7.81           & 7.59           & {\ldots}&$7.50\pm0.05$\\
S\,\textsc{ii}       & 7.15           & 7.21           & {\ldots}&*$7.14\pm0.06$\\
Fe\,\textsc{iii}     & 7.52           & 7.61           & {\ldots}&$7.52\pm0.03$\\
\hline
$M/M_\odot$      & $11.5\pm1.1$ &$10.0\pm1.0$ & $4.2\pm0.5 $\\
$R/R_\odot$      &  $6.5\pm1.5$ & $5.1\pm1.4$  &$2.2^a$ \\
$\log L/L_\odot$ &$4.1\pm0.2$ &$3.9\pm0.2$ &$2.4^a$\\
$M_V$ (mag)   & $-3.2\pm0.5$ &$-2.6\pm0.6$ &$+0.1^a$\\
$\tau$ (Myr)    & $13.5^{+1.6}_{-2.1}$ &$14.0^{+2.2}_{-8.0}$& consistent\\
$\tau/\tau_\mathrm{MS}$ &$0.7\pm0.1$&$0.6\pm0.1$&$<0.1$\\
scale factor     & $0.59\pm0.05$ & $0.37\pm0.03$& $0.04^a$\\\hline %
\end{tabular}\\
$^a$ Adopted from stellar models according to the age of the massive companions.\\
\end{table*}

\begin{figure*}
   \centering
\includegraphics[bb= 2 2 557 737,width = 14cm, height=.95\linewidth,angle=-90]{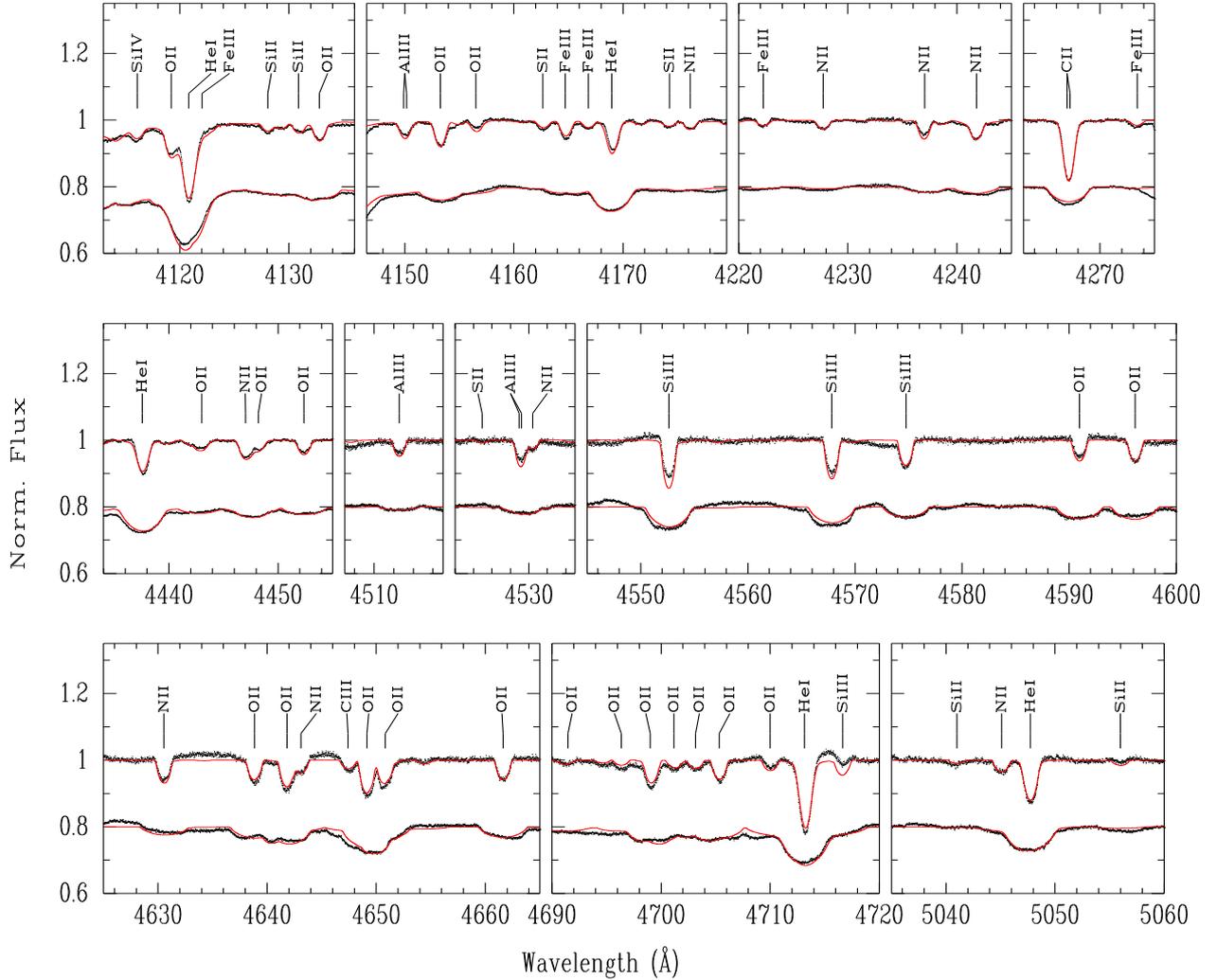} 
    \caption{Synthetic (red) vs. separated spectra (black) of
    components  Ca1 (upper) and Cb (lower set of curves).}         \label{fig:specfit}
   \end{figure*}

\begin{figure*}
   \centering
   \includegraphics[bb= 2 2 550 737, width=10cm,height= 18cm,angle=-90]{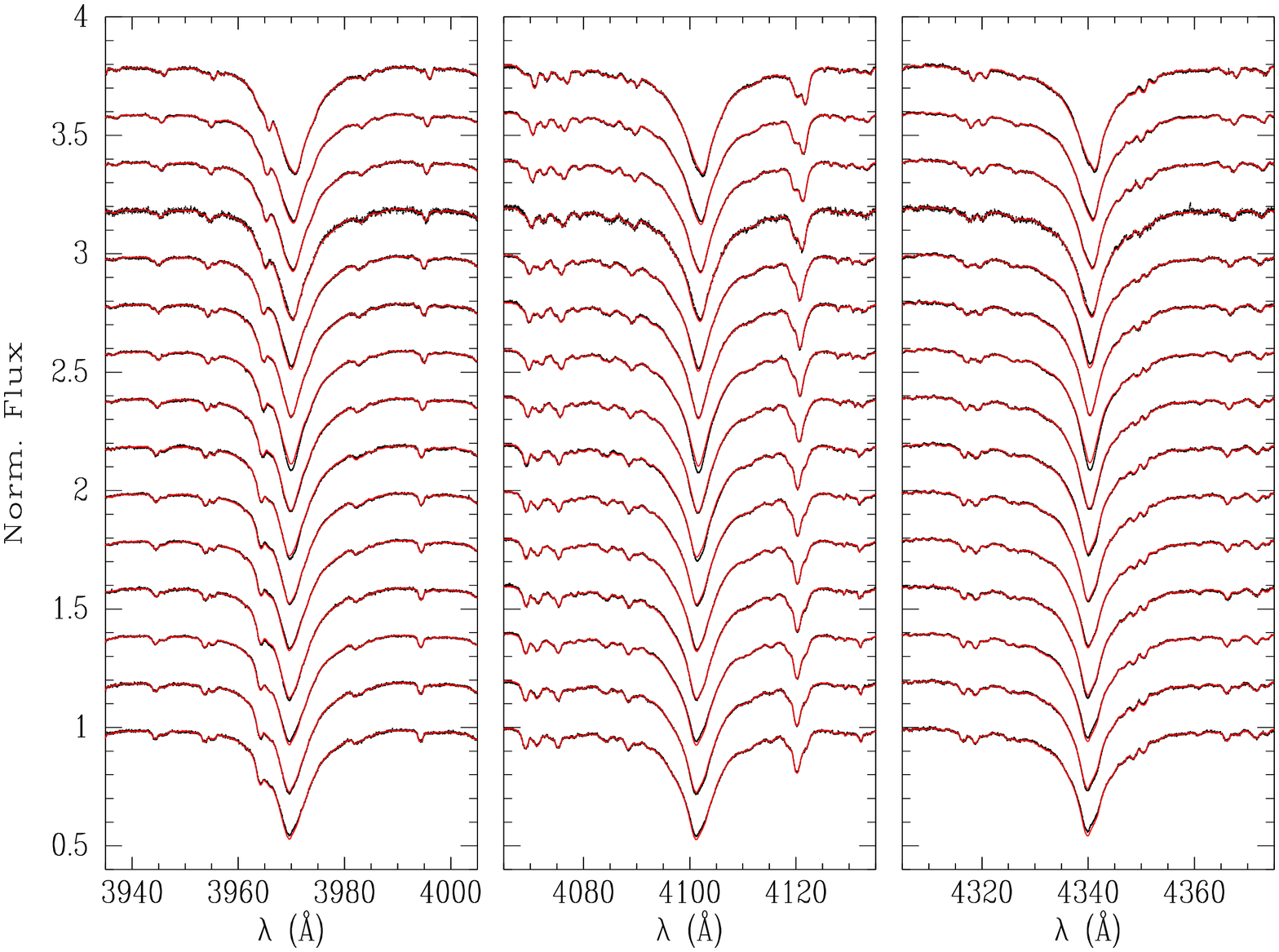}
  \caption{Profiles of several H\,\textsc{i} lines at different orbital phases. 
Comparison of composite synthetic spectra (red line) with UVES spectra (small black dots).}
         \label{fig:HI}  
   \end{figure*}

Grid-based procedures for the analysis of composite spectra
\citep[e.g.][]{2014A&A...565A..63I}
could not be applied without prohibitively large extensions of the
precomputed grids because of the chemically peculiar nature of 
the star Cb (see below). Therefore, the line-profile fitting
was performed on the basis of dedicated microgrid computations
 centred around a set of atmospheric parameters for each star, which were
refined iteratively. The aim was to derive those parameter combinations 
for effective temperature $T_\mathrm{eff}$, (logarithmic) surface gravity
$\log g$, microturbulence $\xi$, projected rotational velocity $v\sin i$, 
(radial-tangential) macroturbulence $\zeta$, helium abundance $y$ (by number), 
and individual metal abundances that facilitate an overall good fit to
be achieved for
{\sc i)} the disentangled spectra (weak metal lines only) and {\sc ii)} the observed composite 
spectra (broad H and He lines, metal lines). The derived parameters
should also facilitate the stars to share a common isochrone and a
common spectroscopic distance, and lead to consistent flux scaling
factors. Many of the parameters are interrelated in the analysis methodology, 
hence the need for an iterative approach.

 In an initial step, we estimated starting values for projected rotational velocities by 
applying to the separated spectra of the three components the technique 
introduced by \citet{2011A&A...531A.143D}.
Initial values for atmospheric parameters were based
on spectral types and considered two additional external constraints:
{\sc i)} the evolutionary age must be the same for all three components, and
{\sc ii)} the mass-ratio $M$(Ca2)/$M$(Ca1) must be consistent with the $q$ value derived from the 
spectroscopic orbit. When all three stars are considered to belong to the same isochrone, 
atmospheric parameters and relative fluxes are no longer independent. 
For fixed temperatures, one single parameter (e.g.\ age) determines $\log g$ and 
relative fluxes for all three stars. In these calculations we used the
stellar model grids by \cite{2012A&A...537A.146E}. This strategy allowed us to adopt starting 
gravity values without having individual hydrogen line profiles.

In a second step, both $T_\mathrm{eff}$ and $\log g$ were refined 
on the basis of  ionisation equilibria,
i.e.~all  ionisation stages of an observed element have to indicate the same abundance 
(within the uncertainties), employing narrow He\,\textsc{i/ii} lines, C\,\textsc{ii/iii},
O\,\textsc{i/ii}, Si\,\textsc{ii-iv}, and S\,\textsc{ii/iii}. 
Individual metal lines were fitted using spectral synthesis.
Analysing the disentangled spectra required a double-iterative procedure,
one on the atmospheric parameters and chemical abundances and another in the 
scaling factors of the disentangled spectra.
New scaling factors were derived from stellar parameters calculated
during the analysis of the separated spectra via metal  ionisation equilibria.
In this instance we relaxed the  strict constraint imposed by the theoretical isochrones.
From this preliminary analysis, we confirmed that Cb is a He-rich star.

 A final check within an iteration step was performed on the observed
composite spectra using the refined atmospheric parameters, chemical
abundances, and scaling factors. Synthetic composite spectra for particular orbital 
phases were built by combining the three synthetic spectra of the component stars 
after shifting by radial velocity, broadening with appropriate 
rotational and macroturbulent profiles, and scaling according to their relative fluxes.
At this stage we were able to include
H\,\textsc{i} and broad He\,\textsc{i} lines in the analysis.
This yielded further estimates for corrections of parameters, for which dedicated synthetic
spectra were again computed. We then returned to the investigation 
of the separated spectra as discussed in the previous paragraph, etc.,
requiring several iteration steps until no further need for changes in the parameters 
was found from the spectroscopic analysis, meeting also the additional
constraints on consistent ages, spectroscopic distances, the Ca2/Ca1
mass ratio and the flux scaling factors. 
The procedure turned out to be challenging and time consuming
due to the large number of variables, further complicated because of 
the chemical peculiarities of Cb.

The finally adopted atmospheric and fundamental stellar parameters, and
the abundances for the components of HD\,164492C are summarised in Table~\ref{tab:atmpar}.
For comparison, chemical abundances as derived from a sample of
chemically normal and single early B-type
stars in the solar neighbourhood \citep{2012A&A...539A.143N} --
the `cosmic abundance standard' (CAS) --, are also given.

Figure~\ref{fig:specfit} shows a selection of spectral lines used in the  abundance determination.
An overall good to excellent match between the model and the disentangled
spectra is obtained.
 Even so, for some lines the residuals are significantly above the noise level, which is low in the
reconstructed spectra (typically 1/400--1/500 of the continuum level). 
These discrepancies might be a sign of surface chemical spots,
although it is also possible that they are remains of the spectral separation process. 
With the velocity amplitude of star Ca1 being
within the width of the line profile of star Cb, the resulting line intensities in the reconstructed spectra of Cb and Ca1 
are strongly correlated. 
  
Despite the high signal-to-noise ratio of the separated spectra of the components Ca1 and Cb, 
the uncertainties in chemical abundances are estimated to be about 0.2\,dex.
The main error source is the scale factor introduced by the relative fluxes and in the case
of Cb the uncertainty in the continuum definition in the region of broad lines or blends.
The fitting of Balmer lines was consistent with atmospheric parameters derived 
from metal lines throughout the orbital period, see Fig.~\ref{fig:HI}.

The star Cb is a He-rich star  ($y$\,=\,0.35$\pm$0.04, i.e.~about
+0.6\,dex above the CAS value) with some metal peculiarities,
particularly moderate overabundance of O and Si, and underabundances of 
C, Ne and Mg. Abundance anomalies are frequently found in
He-strong stars  as a consequence of the different coupling of
individual ions to the stellar wind and due to the formation of
chemical inhomogeneities (spots) because of the magnetic field 
\citep[see e.g.][]{2016A&A...587A...7P}, and Cb is not
an exception. 
The star Ca1 has a chemical composition compatible with the CAS, though by trend a slightly
more metal-rich composition is indicated.

The faint component Ca2 was assessed only through the disentangled spectrum since
its contribution to broad H and He lines is comparatively small making the fitting 
of composite spectra insensitive to the parameters of this star. However, 
its narrow metal lines are well reconstructed in the spectral
disentangling, though the overall signal-to-noise ratio is low
and the uncertainty in its relative flux is large, which affect systematically the 
line intensities. Moreover, without good neutral or doubly  ionised metal lines, 
it was not possible to derive the surface gravity for this star. Instead, we 
adopted a value  consistent with the age of the more massive stars.
As a consequence, we refrain from providing abundance values for Ca2
in Table~\ref{tab:atmpar}. On the other hand, for any reasonable scale
factor the spectrum appears peculiar, the star being apparently a
Si-rich Bp type with its effective temperature ranging 
from  about 14\,000 to 16\,000\,K.

 \section[]{Physical parameters}\label{sec:param}

 \begin{figure}
   \centering
   \includegraphics[bb=2 30 557 737, height=8.8cm,width=7cm,angle=-90]{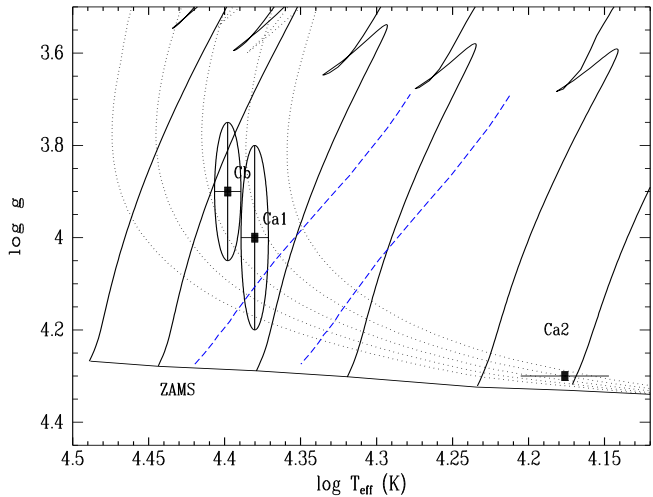} 
   \caption{$\log T_\mathrm{eff}-\log g$ diagram. Isochrones correspond to Geneva stellar models
   for rotating stars with $Z=0.014$. Dotted lines are (from left to right) isochrones for log age 6.9, 7.0, 7.1, 7.2, and 7.3.
   Continuous black lines are (from right to left) stellar tracks for 4, 5, 7, 9, 12, and 15\,$M_\odot$. The strip marked with the two blue dashed lines corresponds to the stellar 
   models of star Ca1 for which the flux-ratio Ca2-to-Ca1 is 0.16$\pm$0.03, asssuming a temperature 15\,000$\pm$1000 K for Ca2. }
   \label{fig:tlogg014}
   \end{figure}

We  fitted the observed spectra assuming that they are formed by three stars of the same age.
Figure~\ref{fig:tlogg014} shows the position of the three companions in the $\log T_\mathrm{eff}-\log g$ diagram.

The adopted parameters are listed in the third block of Table~\ref{tab:atmpar}. 
The spectroscopic $\log g$ value of stars Cb and Ca1 suggests that they are somewhat evolved within the main-sequence, with
an age of  $\sim$12-15\,Myr
according to Geneva models for rotating stars with
metallicity $Z=0.014$.
Using these masses and the radial velocity curve, we estimate the orbital inclination of the Ca1-Ca2 system to be $i=52\pm3^{\circ}$.

For the adopted isochrone, assuming star Ca2 is a normal main-sequence star of the same age,
the relative flux of star Ca2 would be about 0.040$\pm$0.013. 
The  observed  intensity of spectral lines of this star
in the composite spectrum, however, suggests a scaling factor $0.061\pm0.012$.
In other words, the intensity of Ca2 lines in the spectrum suggests that stars Ca1 and Cb
are somewhat less luminous than found in the spectral analysis.
If the  relative contribution of flux for star Ca1 is of the order of 0.37, then the intensity
of the Ca2 lines suggests that the flux ratio Ca2-to-Ca1 is 0.16$\pm$0.03.
The strip between blue lines in  Fig.~\ref{fig:tlogg014} marks the positions of star Ca1
that are consistent with a temperature 15\,000$\pm$1000 K for Ca2 and a flux ratio 0.16$\pm$0.03.
In short, the spectrum of Ca2, under the assumption of being a normal star of the same age,
 suggests the system to be somewhat younger, at most 10\,Myr.

At periastron the components of the Ca pair are at a separation of
about 26\,$R_\odot$, 
only 3-4 times larger that the sum of their radii.
This is not close enough, however, for the orbit and the rotation velocity of the companions
to have suffered significant changes by tidal effects.
Calculations with the binary evolution code
by \citet{2002MNRAS.329..897H} indicate that no significant evolution of the orbit and
rotational velocities would occur until the primary approaches the end of the main-sequence
at about 25\,Myr.
With an eccentricity of $e=0.532$ the pseudo-synchronous rotational period would be 3.994\,days. 
It is interesting that the projected rotation corresponding to the 
pseudo-synchronous regime, as calculated from the adopted stellar parameters,  $51\pm9$\,km\,s$^{-1}$, is in agreement with the observed $v\sin i$ value. 
In other words, the primary Ca1 seems to be rotating close to the pseudo-synchronous regime,
although this  would not be caused by  tidal forces.

The stellar parameters determined for the components of this multiple system offer the opportunity to 
calculate the spectroscopic distance modulus.
Even though various works have determined the distance to the Trifid Nebula cluster using other 
stars or properties of the nebula, this issue is still controversial. 
The high and variable extinction, and mainly the anomalous extinction
law has contributed to this. 
\citet{1985ApJ...288..164L} reported an extinction law with $R_V=5.1$ and
estimated a distance of 1.67\,kpc.
\citet{2011A&A...527A.141C} confirmed this high $R_V$ value, but obtain
a significantly larger distance.
From the  colour distribution of stars along the line-of-sight to strongly absorbed
regions within the Trifid nebula, they derived the extinction distribution 
along the line-of-sight and calculated a distance of $2.7\pm0.5$\,kpc to the nebula.
A similar value had been obtained from $UBV$ photometry of stars HD\,164492A, B, C, and E, by \citet{1999A&AS..134..129K}.

Unfortunately none of the stellar components of the multiple system HD164492  is
included in the {\it Gaia} first data release \citep{gaia}.
Parallaxes are available in this catalogue for a few other probable members of the cluster NGC\,6514
but the uncertainties are too large ($\sim$50\%) for a reliable determination of the distance.

The adopted stellar parameters for the three stars studied in the present paper  correspond %
to a total absolute visual magnitude of $-3.72\pm0.53$ mag.
The apparent $V$ magnitude of this triple system is not well known.
The photometric measurements given by  \citet{1999A&AS..134..129K} involve
also the component HD\,164492D. 
Using the light ratio estimated by the same authors (3.9 at $\lambda=514$\,nm), we obtain: $V=8.91$\,mag for 
the integrated magnitude of the system HD\,164492C.
Assuming a reddening $E(B-V)= 0.33\pm0.03$\,mag \citep[][for stars A, B, and C]{1999A&AS..134..129K},
and $R=5.1$, we derive a distance of $1.55^{+0.45}_{-0.35}$\,kpc,
supporting the results of \citet{1985ApJ...288..164L}.

\section{Spectral variability}

\begin{figure}
   \centering
   \includegraphics[bb= 2 20 550 497, width=7.5cm,height=8cm,angle=-90]{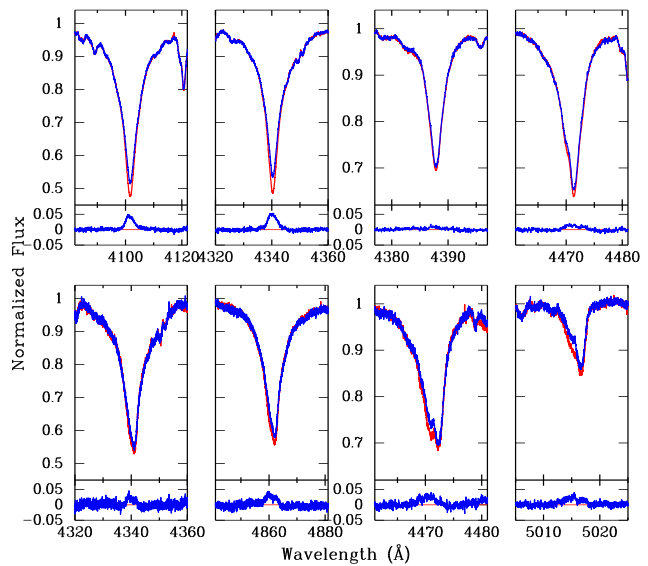}
  \caption{Profile variations of  H\,\textsc{i} and He\,\textsc{i} lines.
  Upper panels: comparison of UVES spectra at HJD~2\,456\,869.5077 (red) and HJD~2\,456\,919.5607 (blue).  
  Lower panels: comparison of HARPS spectra at HJD~2\,457\,091.8426 (red) and HJD~2\,457\,317.5185 (blue). In the lower part of each panel, the difference between the two plotted spectra is shown. }\label{fig:HHe}  
   \end{figure}
  
Even though the combination of the reconstructed spectra of the three components 
reproduces  satisfactorily the observed spectra, particularly the metallic lines,
some observations show small differences in the core of Balmer lines  (see Fig.~\ref{fig:HI} and \ref{fig:HHe}).
In the spectra with strong H lines some strong He lines appear also enhanced, although these
variations are always below 1\% of the continuum and should be considered as marginal. 
 These line-profile variations are not correlated with the orbital phase of the system Ca1-Ca2.
This is shown in Fig.~ \ref{fig:HHe} where we compare two UVES spectra (HJD 2\,456\,869.5077 and 2\,456\,919.5607) taken at similar orbital phase (near conjunction) but presenting different 
intensity in H lines and some He lines. Two HARPS spectra exhibiting differences in H and He lines are also shown. In this case the observations were taken on
 HJD 2\,457\,091.8426 and 2\,457\,317.5185, which corresponds to phase $\sim$0.09
(near quadrature with star Ca1 shifted to the red). 
In some cases these variations are noticeable on consecutive nights, suggesting a short
period that  might be related with a non-uniform surface of the rotating star Cb.
In most of the spectra, however the spectral variations are comparable to the spectrum noise or 
the continuum uncertainties, making difficult its use for deriving the rotation period.

More than one mechanism could be responsible for these spectral variations.
One possible explanation is a temperature variation over the surface of this star.
In the temperature range of star Cb, a drop in temperature would strengthen both He and H lines.
In chemical peculiar stars, the regions with a great concentration of some particular elements (chemical spots) present a higher opacity,  which increases the temperature due to backwarming
(see \citealp{2007A&A...470.1089K} for a clear example in a He-strong star).
On the other hand, in massive magnetic stars, temperature variations can be related to bright spots 
that arise due to the lower density (lower optical thickness) of the gas in surface regions with higher magnetic pressure \citep{2011A&A...534A.140C}.

\begin{figure}
   \centering
\includegraphics[height=7cm,width=8cm]{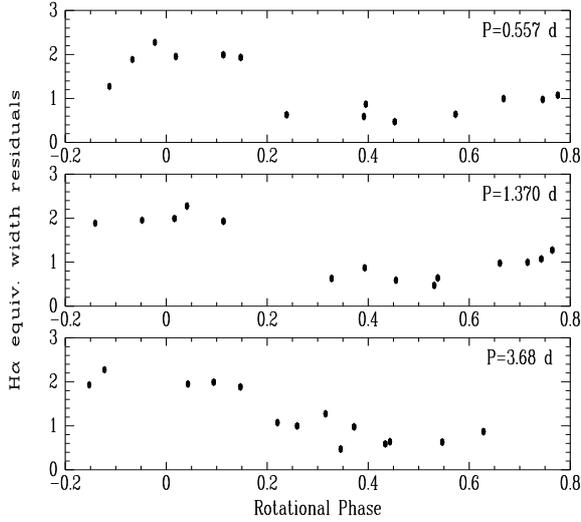} 
    \caption{Flux excess in H$\alpha$ phased with three probable periods.}   \label{fig:Ha}
   \end{figure}

Besides the mentioned variations in the core of Balmer lines, the line H$\alpha$ shows variable wings
which in a few spectra appear in emission. 
To analyse these variations we subtracted from the UVES spectra synthetic spectra of the three components as 
calculated in Sect.~\ref{sec:abun}.
These difference spectra were used to calculate residual equivalent widths integrating in 
a wavelength window of 80 \AA \ around H$\alpha$.
We find that all UVES spectra exhibit an H$\alpha$ excess in the range 0.5--2.3 \AA \ over the synthetic model, except
for the spectrum taken at HJD 2\,456\,779.9, which has a much stonger emission (3.6 \AA), probably due to the 
contamination by the Herbig Be star (HD\,164492D) in the vicinity of HD\,164492C. 
A period search on the sample of fourteen UVES difference spectra shows that these variations might present
a periodic behaviour, although the period is not unambiguously determined with the available 
UVES observations.
The most probable periods are 0.557 d, 1.370 d, and 3.68 d.
Figure~\ref{fig:Ha} shows the equivalent width curves built with these three periods.
The estimated uncertainties are of the order of 0.05 \AA. Since FEROS and HARPS are  fibre-fed 
instruments, it is difficult to take into account the background variations. As a results, we do not detect in the
spectra obtained with these instruments clear emission variations similar to those detected in UVES spectra.

\section{Magnetic fields}\label{sec:magfield}

\begin{table}
\caption{Logbook of the HARPS spectropolarimetric observations of HD\,164492C.
Column 1: Heliocentric Julian Date; 
Column 2: Orbital phase of the subsystem Ca1-Ca2; 
Column 3: Rotational phase of star Cb   using P=1.59689 d and T$_0$=2\,457\,001.263;
Column 4: Longitudinal magnetic field measured using metallic and He lines;
Column 5: Longitudinal magnetic field measured using only metallic lines.}
\label{tab:mag}
\centering
\begin{tabular}{lccr@{$\pm$}rr@{$\pm$}r}
\hline
\multicolumn{1}{c}{HJD} &
\multicolumn{1}{c}{Orbital} &
\multicolumn{1}{c}{Rotational} &
\multicolumn{2}{c}{$\left< B_{\rm z}\right>_{\rm all}$} &
\multicolumn{2}{c}{$\left< B_{\rm z}\right>_{\rm no\,He}$} \\
 &
 Phase &
 Phase &
 \multicolumn{2}{c}{[G]} &
 \multicolumn{2}{c}{[G]}  \\
 \hline
2\,456\,445.7946 &0.5559	&	0.1548	& 615  & 22 &   309 & 64\\
2\,456\,446.7934 &0.6355	&	0.7803	& 351  & 24 &   203 & 65\\
2\,456\,770.7700 &0.4808	&	0.6606	& $-$52& 16 & $-$78 & 41\\
2\,456\,771.7850 &0.5618	&	0.2963&  11  & 15 &   18  & 45\\
2\,457\,091.8426 &0.0944	&	0.7225	& $-$72& 27 & $-$27 & 83\\
2\,457\,092.8282 &0.1730	&	0.3397	&   48 & 26 & $-$89 & 77\\
2\,457\,093.8478 &0.2544	&	0.9782	& 404  & 22 & 419   & 56\\
2\,457\,094.8153 &0.3316	&	0.5840	& 325  & 32 & 42    & 67\\
2\,457\,178.6948 &0.0231	&	0.1110	& 784  & 20 & 467   & 37\\
2\,457\,179.7426 &0.1067	&	0.7671	&$-$46 & 36 & 228   & 76\\
2\,457\,317.5185 &0.0977	&   0.0450	& 905  & 37 & 406   & 62\\
\hline
\end{tabular}
\end{table}

 \begin{figure*}
   \centering
   \includegraphics[height=20cm,width=17cm,angle=0]{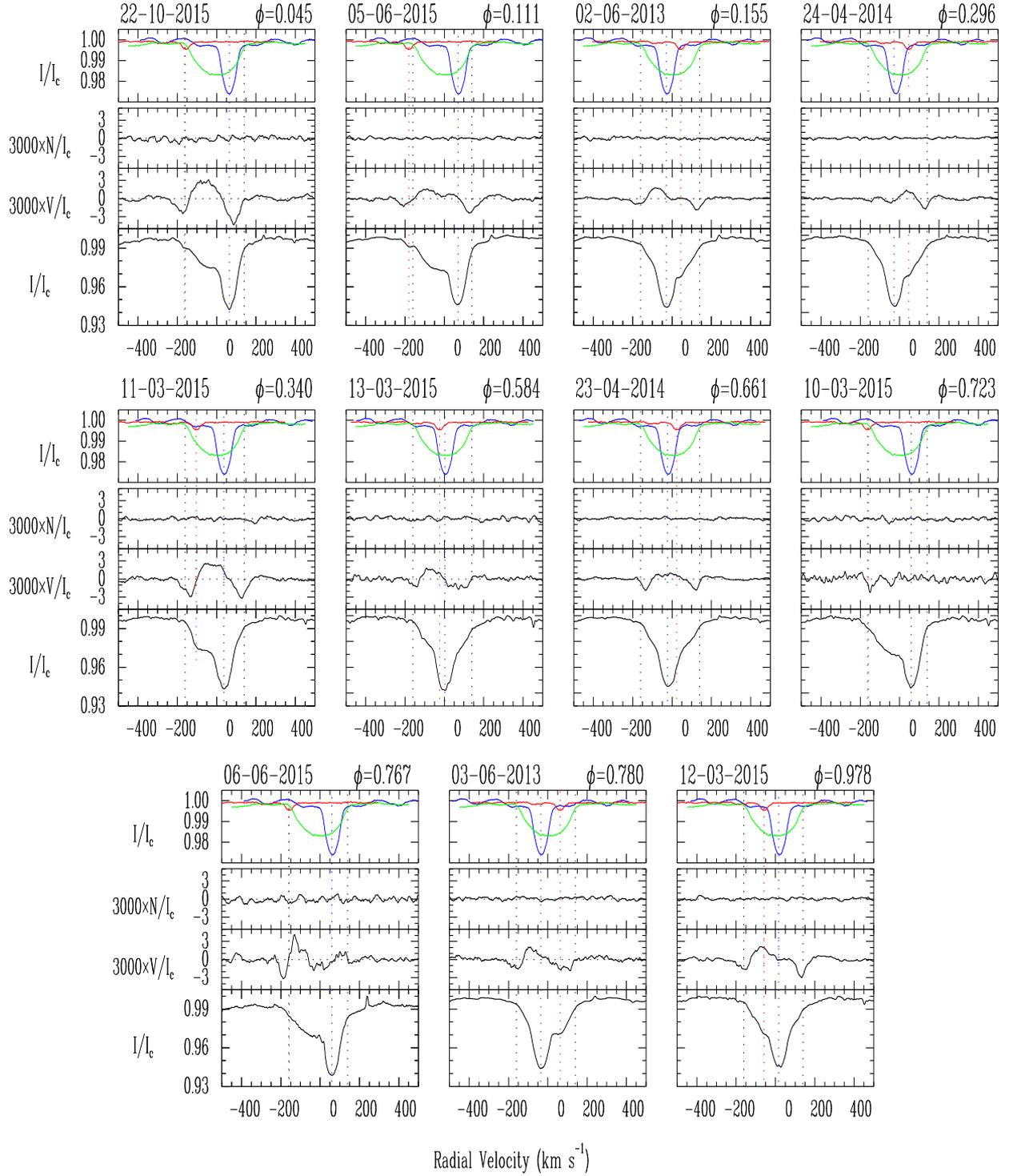} 
   \caption{$I$, $V$, and $N$ SVD profiles  obtained for the system HD\,164492C on eleven different epochs
using a line mask including all lines. In the upper panel we present the SVD Stokes~$I$ line profiles for each component at different
positions in velocity space according to their radial velocity shifts due to orbital movement.
The $V$ and $N$ profiles were expanded by 
the indicated factor for better visibility. Note that 
the plots are ordered according to rotational phase and not to increasing Julian date.}
         \label{fig:magf_1}
   \end{figure*}

\begin{figure*}
   \centering
   \includegraphics[height=20cm,width=17cm,angle=0]{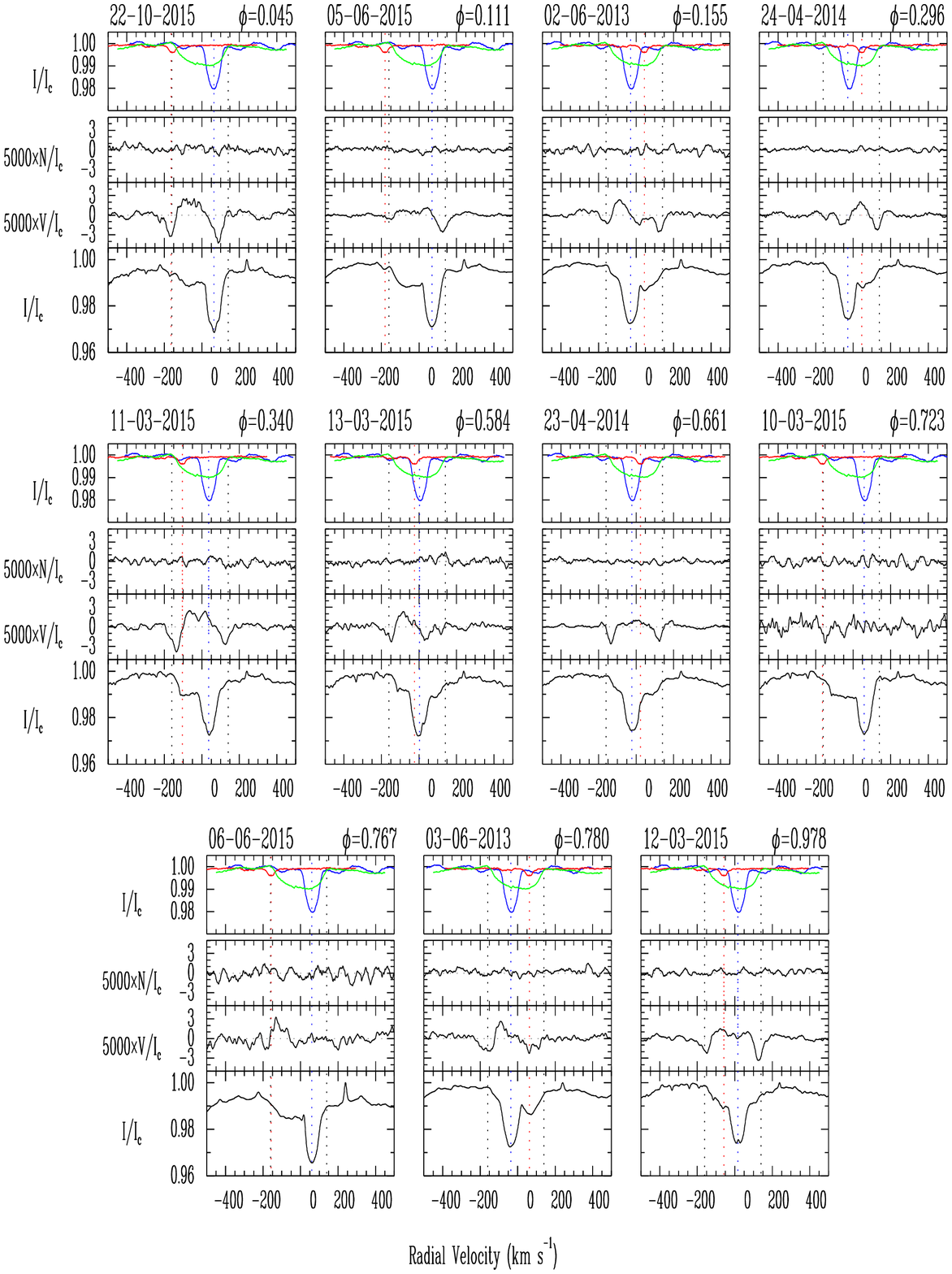} 
   \caption{The same as in Fig.~\ref{fig:magf_1} but using exclusively metal lines. }
       \label{fig:magf_2}
\end{figure*}

The reduction and magnetic field measurements were carried out using dedicated software packages developed for the 
treatment of high-resolution spectropolarimetric data.
The spectrum reduction and calibration was performed using the HARPS data reduction pipeline available at the ESO 3.6-m telescope in Chile. 
The normalisation of the spectra to the continuum level consisted of several steps described in detail by \citet{2013AN....334.1093H}. 
The software package used to study the magnetic field in HD\,164492C, the so-called  
multi-line Singular Value Decomposition (SVD) method 
for Stokes profile reconstruction was recently introduced by \citet{2012A&A...548A..95C}. 
More details on the basic idea of SVD can be found in \citet[][]{2009IAUS..259..633C}.
The results of our magnetic field measurements, those for the line list including all lines and the line list excluding 
He lines, are presented in 
Table~\ref{tab:mag}, where we also collect information about the heliocentric Julian date 
for the middle of the exposure, and the rotational phase assuming a period of 1.5969\,d (see below).

    \begin{figure}
   \centering
  \includegraphics[height=5.5cm,angle=0]{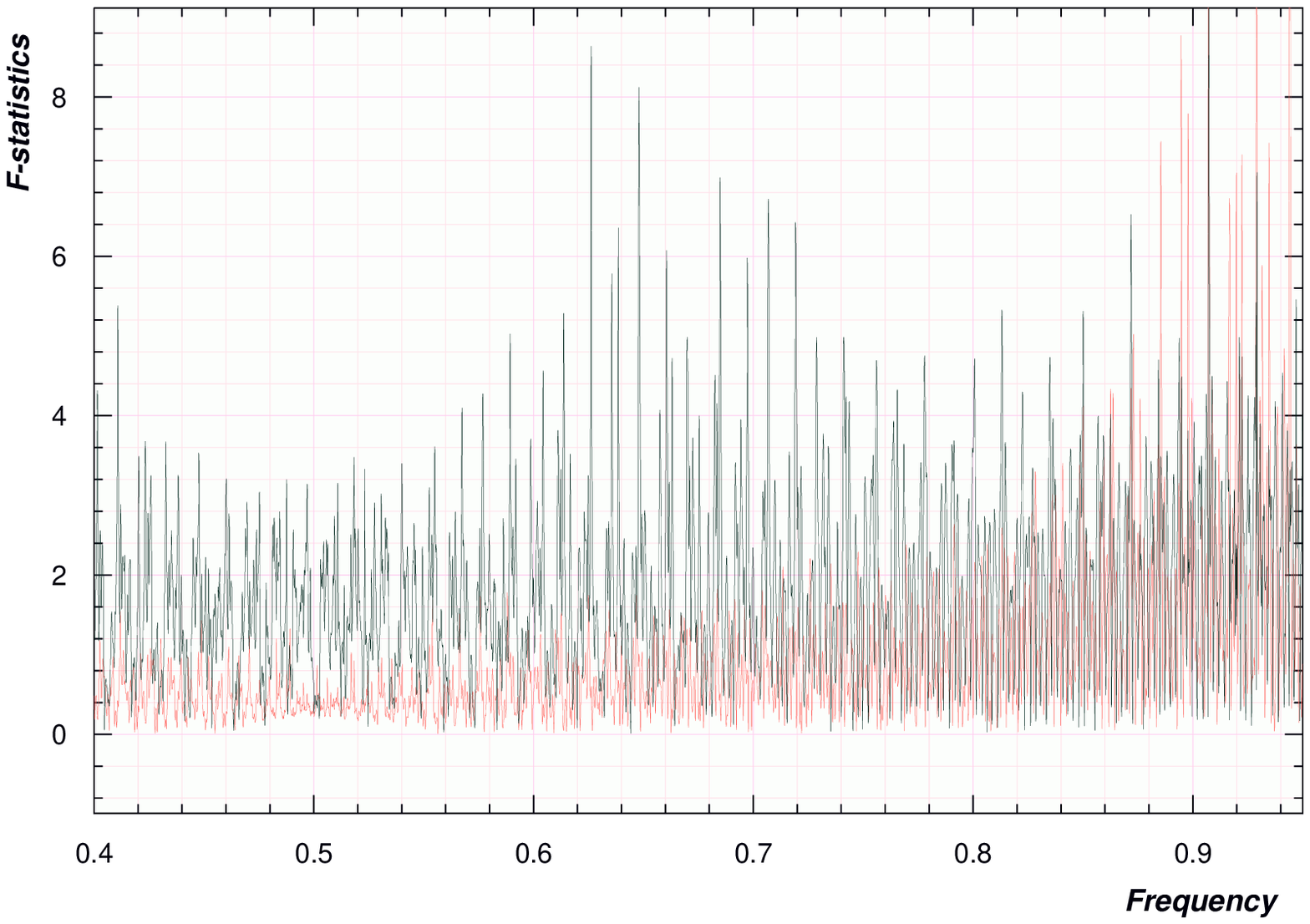} %
   \caption{Periodogram for the mean longitudinal magnetic field measured without He lines. 
   The peak at 0.6262 d$^{-1}$ corresponds to a period of 1.5969\,d.
   The window function is plotted in red.} 
         \label{fig:periodogram}
   \end{figure}

The resulting mean Stokes~$I$, Stokes~$V$, and Null profiles obtained by using the SVD
method with 109 metallic lines in the line mask and  with the line mask including He lines (125 lines) are presented in 
Figs.~\ref{fig:magf_1} and \ref{fig:magf_2}. Null polarisation spectra were calculated by combining the subexposures in such a way that 
polarisation cancels out. The line mask was constructed using the VALD database 
\citep[e.g.\ ][]{kupka2000}. 
The mean longitudinal magnetic field was estimated from the SVD reconstructed Stokes~$V$ and $I$
using the  centre-of-gravity method (see e.g.\ \citealt{CarrollStrassmeier2014}).
 The velocity window used in these measurement was about 400 \kms{} wide, which includes all three components in all phases. 
We note that due to the fact that HD\,164492C is a hierarchical triple system and all three components remain blended in 
all observed spectra, the estimation of the magnetic field can only be done assuming that HD\,164492C is a single star.
A definite detection of a mean longitudinal magnetic field, $\left<B_{\rm z}\right>$, with
a false alarm probability (FAP) smaller than $10^{-6}$,
was achieved on all epochs apart from the observation on HJD 2\,457\,091.8426 (orbital phase 0.09) where a marginal detection was 
achieved for the line mask including all lines, and no detection for the line mask including exclusively
metal lines. 
The analysis of the diagnostic Null profiles showed non-detections at all phases apart from the same 
epoch at the orbital phase 0.09 where we achieve a marginal detection with FAP of about $2\times10^{-5}$
for the same line mask with metal lines.

In the framework of our ESO Large Programme 191.D-0255, the first spectropolarimetric observations of HD\,164492C were carried 
out in 2013 with the 
low-resolution FORS\,2 spectrograph and were complemented in the same year by two observations using HARPS in 
spectropolarimetric mode \citep{2014A&A...564L..10H}. While FORS\,2 observations indicated a strong longitudinal 
magnetic field of about 500--600\,G,
the HARPS observations revealed that HD\,164492C could not be considered a single star as it possessed one or two companions.
Given the complex configuration and shape of the Stokes profiles, it was also impossible
to conclude exactly which components possessed a magnetic field.
The detection of a significant Stokes~$V$ signature in the SVD and LSD profiles at about $-100$\,km\,s$^{-1}$ and 
$+150$\,km\,s$^{-1}$ from the line core of the primary suggested that
more than one component might hold a magnetic field. On the other hand, assuming that HD\,164492C is a single star, we obtained
results very similar to those obtained 
with FORS\,2: between 500 and 700\,G for the first HARPS observing night, and between 400 and 600\,G for the second HARPS night.

  \begin{figure}
   \centering
\includegraphics[width=6cm,angle=-90]{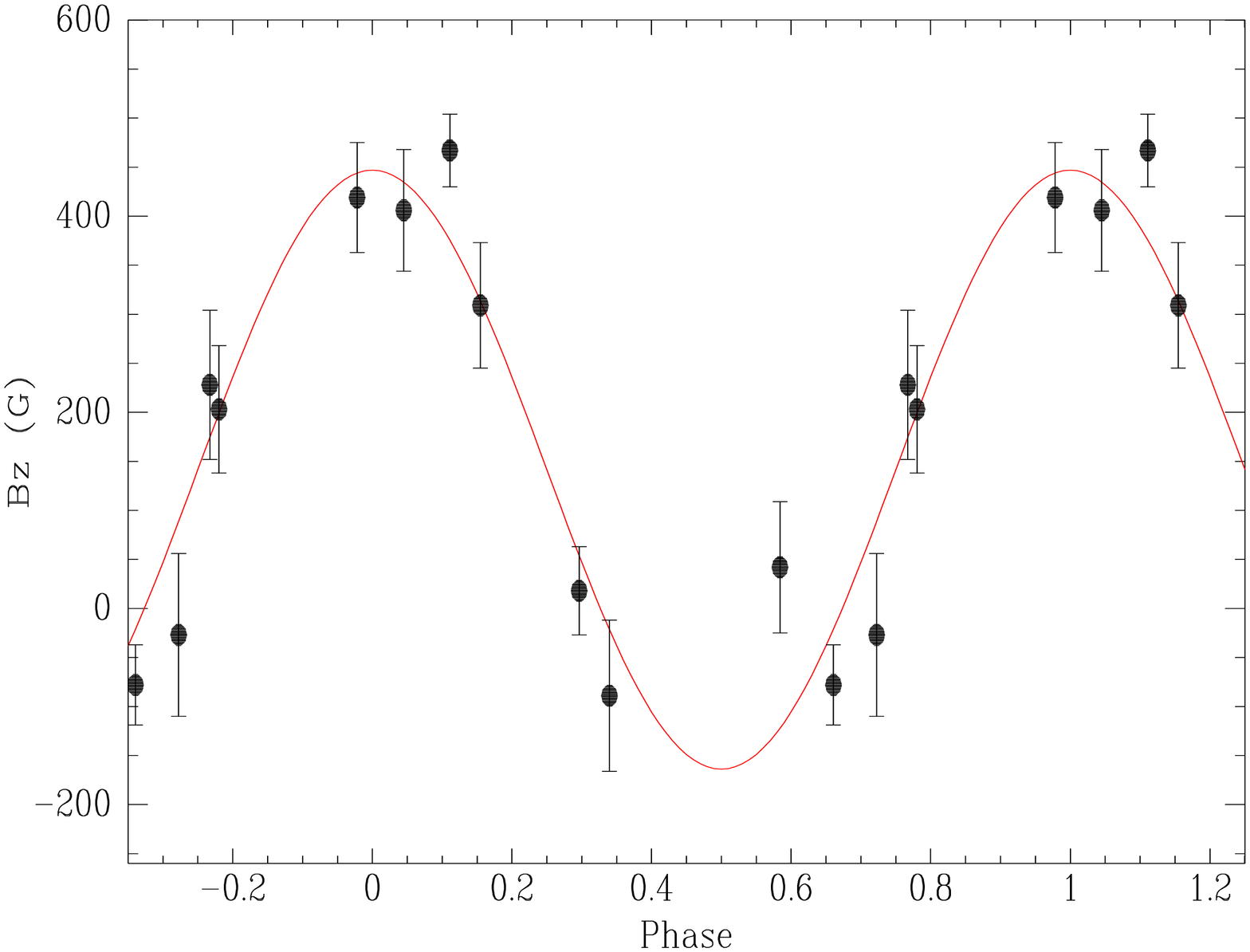} %
   \caption{Magnetic field curve calculated with a period of 1.5969\,days. The plotted values    correspond to the longitudinal magnetic field measured using metallic lines. }
         \label{fig:magcurve}
   \end{figure}

Since 2013 we were able to acquire nine additional HARPS spectropolarimetric observations.
Admittedly, even having at our disposal observations at eleven epochs, the interpretation of the temporal behaviour 
of the Stokes~$V$ profiles is a challenging task because the line profiles of the three stars overlap in all
orbital phases. In fact, the radial velocity amplitude of the star Ca1, the primary of the
spectroscopic binary, is smaller than the
projected rotational velocity of the star Cb, which has a similar spectral type. The radial velocity amplitude of the 
component Ca2 is larger, but still overlaps the wings of the SVD line profile of the rapidly rotating component Cb.
We note, however, that at several epochs (for example on 02-06-2013 and 24-04-2014,
corresponding both to the orbital phase 0.56) we observe
significant features in the Stokes~$V$ profiles that are located in the velocity space far from the positions 
corresponding to stars Ca1 and Ca2. These features can only be attributed to the star Cb. 
It is possible therefore, to identify the spectroscopic component Cb as a star possessing a magnetic field.
This is in line with it being a He-rich star. The group of He-peculiar stars are as a rule strongly magnetic \citep{1979ApJ...228..809B}. %
Given the rather fast changes in the shape of Zeeman features over consecutive nights, we expect that the rotation period of this component 
is rather short.
It is obvious that the magnetic character of the star Ca1 cannot be addressed as long as the magnetic variations
of the star Cb are not modelled. Still, due to the severe distortion of the Zeeman features at the positions corresponding to the red or blue
wing of the Stokes~$I$ profile of the component Cb at some epochs, we cannot
exclude that also the star Ca1 possesses a magnetic field. Indeed, sometimes only the 
blue part of the Zeeman feature is visible (for example at HJD 2\,457\,091.8426 %
- rotational phase 0.72, or HJD 2\,457\,179.7426 %
- rotational phase 0.77)  and sometimes only the red part of the Zeeman feature 
(for example at HJD 2\,456\,771.7850 %
- rotational phase 0.30).
In the working hypothesis assuming that Cb is the only magnetic star and shows a  
dipole structure, 
for all observational aspects of the dipole configuration, i.e.\ either close to the magnetic poles or 
close to the magnetic equator, we would always expect a certain symmetry in the appearance of the Zeeman features relative to the
line  centre of the component Cb, unless strong contrast chemical inhomogeneities are present on the surface. 
Interestingly, since He is especially strong in the component Cb, Zeeman features appear less noisy in the 
Stokes~$V$ spectra obtained including He lines at the orbital phases mentioned above.
As for the star Ca2, we can conclude that its low relative
contribution to the total flux (5-7\% of the total light) makes
its potential magnetic features (because of its probable BpSi type) 
to be virtually undetectable.

From the estimated stellar parameters and the projected rotational velocities,  upper limits for
the rotational period of the stars can be obtained. 
In the case of the star Cb, assuming a radius of $6.3\,R_\odot$ leads to a period of less than about 2.3\,days.

In order to search for rotational modulations of the magnetic field, we performed a
periodicity analysis of the mean longitudinal magnetic field measurements.
Figure~\ref{fig:periodogram} shows the periodogram corresponding to measurements
obtained without the He lines. 
A few probable periods appear in the range 1.4--1.6\,days (frequencies of $0.62-0.72$\,d$^{-1}$).
The highest peak corresponds to $P=1.5969$\,d.
Figure~\ref{fig:magcurve} shows the magnetic measurements using metallic lines phased
with this period.

On the other hand, as mentioned in Sect.~7, the line profiles of H$\alpha$ in UVES spectra exhibit 
excess of emission indicating periodicity with the probable
periods of 0.557 d, 1.370 d, and 3.68 d. 
It is possible that the appearance of the broad emission profiles is related to the presence 
of the magnetosphere around the He-rich Cb component where stellar wind is 
forced to corotate with the star up to the Alfv\'en radius.
Thus, the periodic variability of the emission in the H$\alpha$ line profile can be probably related 
to the rotational period of this component.
\begin{figure}
   \centering
\includegraphics[bb=2 20 557 737,width=6.3cm,angle=-90]{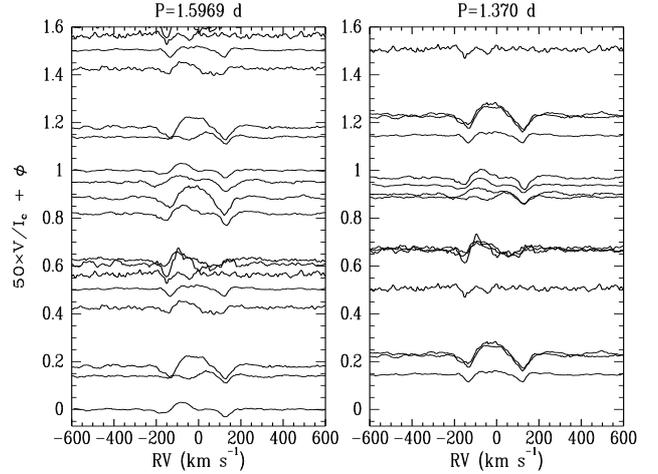} %
   \caption{ Variation of the Stokes~$V$ LSD profiles of HD~164492C (all three components included) with respect to the
   rotational phases calculated with $P=1.370$\,d and $P=1.5969$\,d. In each panel, the $V/I_c$ profiles are scaled by a factor 50
   and shifted according to the rotational phase $\phi$. }
         \label{fig:LSDprofiles}
   \end{figure}
In Figure~\ref{fig:LSDprofiles} we present the Stokes~$V$ profiles obtained for the line mask including 
all lines ordered according to rotational phases calculated with periods 1.370 and 1.597\,days.
The apparent resemblance of Stokes~$V$ Zeeman profiles in observations obtained in similar 
phases suggests that the period of 1.37\,day can also be considered as a tentative rotational period
of the component Cb.
 Our magnetic field measurements, however, do not show a coherent curve if this period is used to phase the data.
We cannot, therefore, give a definite conclusion about the rotational period with the present observational material.

\section{Conclusions}\label{sec:disc}

HD\,164492C is a hierarchical triple system, which on its own is probably a member
of a larger multiple system. 
Two early-B type stars of similar masses (Ca1 and Cb) are bound in a wide binary
system with a separation of the order of 100\,AU \citep{2016BAAA...58..105G}.
One of these stars (star Ca1) has a close companion in an orbit with a period of 12.5\,d.
The radial velocity of the tertiary star Cb coincides within uncertainties with
the  centre-of-mass velocity of the pair Ca12 and with the velocity of the nebula.
The velocity variations due to the orbital motion of the pair Cb-Ca are not detectable,
since with the estimated masses and projected separation, the period would be at least 
100 times longer than the timespan of our observations. 

Star Cb is the most massive star in the system with a mass of about 11\,$M_\odot$. 
It is the apparently fastest rotator of the system and its surface gravity suggests that it is close to the middle
 of its main-sequence lifetime. It is chemically peculiar, exhibiting a strong overabundance of helium
($y$\,=\,0.35, or about +0.6\,dex over the CAS value).
The spectroscopic pair is formed by a 10\,$M_\odot$ primary and a  4\,$M_\odot$ secondary
in an eccentric 12.5\,d orbit. The least massive star is probably a BpSi star.

The distance to the Trifid nebula is not well known and previous determinations range
from 1.7\,kpc %
to 2.7\,kpc. %
We have derived the spectroscopic distance modulus for the triple star HD\,164492C,
member of the Trifid nebula stellar association, and obtained
 a value of 1.5\,kpc, which clearly supports the short distance scale.
Similar results are found from the spectroscopic distance modulus of 
other nearby stars (Przybilla et al., in preparation).  The spectral
morphology is consistent with three main-sequence stars of at least 10\,Myr.
The dynamical age of the Trifid nebula is about 0.3 Myr \citep{1998Sci...282..462C},
while the most accepted age for the first-generation stars,
to which the central O-star HD~164492A belongs, is of the order of 1 Myr \citep{2011ApJ...738...46T}.
In a future paper (Przybilla et al., in preparation) we will address the problem of
the age of the stellar population of M20 in further detail, but we mention here that
the spectral analysis of several stars in the region would suggest that
besides the young population, with spectral morphology consistent with 1 Myr,
there are a few other stars that might be on the same isochrone as HD~164492C.

We confirm the presence of a magnetic field in the triple system HD\,164492C 
and  identified the He-strong tertiary companion Cb as being the magnetic star. 
The detection of the magnetic field in this
chemically peculiar star confirms the hypothesis that He-strong
stars are a hot extension of the Ap phenomenon \citep[e.g.][]{1979ApJ...228..809B}.
However, the observed asymmetry of Zeeman features indicates that the magnetic field cannot be described exclusively by 
a rotating dipole structure on the surface of the Cb component: either the star Ca1 is also magnetic or the
geometry of the magnetic field of the star Cb is relatively complex.
 Our variability analyses of the longitudinal magnetic field and the behaviour of the emission excess
in the H$\alpha$ lines suggest tentative
magnetic periods of 1.6\,d and 1.4\,d. These values, however, should be confirmed by additional observations.

 While this paper was being refereed, \citet{2016arXiv161002585W} published a study
on the same object. This work makes use of most of the observations presented in this 
paper along with an additional series of spectropolarimetric observations obtained
with ESPaDOnS at the Canada-France-Hawaii Telescope.
The spectroscopic orbit obtained by these authors agrees with the one presented here.
Their estimated stellar parameters are also compatible, although their parameters are not 
calculated  solely from the observational data, but through an 
assumption of distance and age of the Trifid Nebula\footnote{Their
apparent distance modulus estimation is rather rough, since they assumed no interstellar 
absorption and derived the distance from the parallax of two other stars in the Trifid Nebula cluster,
one of which (HD\,174492) is clearly not a member of the association.}.
Based on 23 magnetic field measurements they derive a rotational period of 1.3699 d
and obtained inclination and obliquity under the assumption of the dipolar oblique rotator model.

\section*{Acknowledgments}
This work was partially supported by a grant from FONCyT-UNSJ PICTO-2009-0125.
TM acknowledges financial support from Belspo for contract PRODEX GAIA-DPAC.
 MFN acknowledges support by the Austrian Science Fund through the Lise Meitner  programme N-1868-NBL. AK thanks RFBR grant 16-02-00604 А for the financial support. RB acknowledges support from FONDECYT Regular Project 1140076.

\bibliographystyle{mnras}    
\bibliography{bibliomn}

\begin{thebibliography}{}
\makeatletter
\relax
\def\mn@urlcharsother{\let\do\@makeother \do\$\do\&\do\#\do\^\do\_\do\%\do\~}
\def\mn@doi{\begingroup\mn@urlcharsother \@ifnextchar [ {\mn@doi@}
  {\mn@doi@[]}}
\def\mn@doi@[#1]#2{\def\@tempa{#1}\ifx\@tempa\@empty \href
  {http://dx.doi.org/#2} {doi:#2}\else \href {http://dx.doi.org/#2} {#1}\fi
  \endgroup}
\def\mn@eprint#1#2{\mn@eprint@#1:#2::\@nil}
\def\mn@eprint@arXiv#1{\href {http://arxiv.org/abs/#1} {{\tt arXiv:#1}}}
\def\mn@eprint@dblp#1{\href {http://dblp.uni-trier.de/rec/bibtex/#1.xml}
  {dblp:#1}}
\def\mn@eprint@#1:#2:#3:#4\@nil{\def\@tempa {#1}\def\@tempb {#2}\def\@tempc
  {#3}\ifx \@tempc \@empty \let \@tempc \@tempb \let \@tempb \@tempa \fi \ifx
  \@tempb \@empty \def\@tempb {arXiv}\fi \@ifundefined
  {mn@eprint@\@tempb}{\@tempb:\@tempc}{\expandafter \expandafter \csname
  mn@eprint@\@tempb\endcsname \expandafter{\@tempc}}}

\bibitem[\protect\citeauthoryear{{Becker}}{{Becker}}{1998}]{1998ASPC..131..137B}
{Becker} S.~R.,  1998, in {Howarth} I.,  ed.,  Astronomical Society of the
  Pacific Conference Series Vol. 131, Properties of Hot Luminous Stars. p.~137

\bibitem[\protect\citeauthoryear{{Becker} \& {Butler}}{{Becker} \&
  {Butler}}{1988}]{1988AA...201..232B}
{Becker} S.~R.,  {Butler} K.,  1988, \aap, \href
  {http://adsabs.harvard.edu/abs/1988A%26A...201..232B} {201, 232}

\bibitem[\protect\citeauthoryear{{Borra} \& {Landstreet}}{{Borra} \&
  {Landstreet}}{1979}]{1979ApJ...228..809B}
{Borra} E.~F.,  {Landstreet} J.~D.,  1979, \mn@doi [\apj] {10.1086/156907},
  \href {http://adsabs.harvard.edu/abs/1979ApJ...228..809B} {228, 809}

\bibitem[\protect\citeauthoryear{{Borra}, {Landstreet}  \& {Mestel}}{{Borra}
  et~al.}{1982}]{1982ARA&A..20..191B}
{Borra} E.~F.,  {Landstreet} J.~D.,   {Mestel} L.,  1982, \mn@doi [\araa]
  {10.1146/annurev.aa.20.090182.001203}, \href
  {http://adsabs.harvard.edu/abs/1982ARA%26A..20..191B} {20, 191}

\bibitem[\protect\citeauthoryear{{Butler} \& {Giddings}}{{Butler} \&
  {Giddings}}{1985}]{but_gid85}
{Butler} K.,  {Giddings} J.~R.,  1985, Newsletter of Analysis of Astronomical
  Spectra, 9

\bibitem[\protect\citeauthoryear{{Cambr{\'e}sy}, {Rho}, {Marshall}  \&
  {Reach}}{{Cambr{\'e}sy} et~al.}{2011}]{2011A&A...527A.141C}
{Cambr{\'e}sy} L.,  {Rho} J.,  {Marshall} D.~J.,   {Reach} W.~T.,  2011,
  \mn@doi [\aap] {10.1051/0004-6361/201015863}, \href
  {http://adsabs.harvard.edu/abs/2011A%26A...527A.141C} {527, A141}

\bibitem[\protect\citeauthoryear{{Cantiello} \& {Braithwaite}}{{Cantiello} \&
  {Braithwaite}}{2011}]{2011A&A...534A.140C}
{Cantiello} M.,  {Braithwaite} J.,  2011, \mn@doi [\aap]
  {10.1051/0004-6361/201117512}, \href
  {http://adsabs.harvard.edu/abs/2011A%26A...534A.140C} {534, A140}

\bibitem[\protect\citeauthoryear{{Carroll} \& {Strassmeier}}{{Carroll} \&
  {Strassmeier}}{2014}]{CarrollStrassmeier2014}
{Carroll} T.~A.,  {Strassmeier} K.~G.,  2014, \mn@doi [\aap]
  {10.1051/0004-6361/201322825}, \href
  {http://adsabs.harvard.edu/abs/2014A%26A...563A..56C} {563, A56}

\bibitem[\protect\citeauthoryear{{Carroll}, {Kopf}, {Strassmeier}  \&
  {Ilyin}}{{Carroll} et~al.}{2009}]{2009IAUS..259..633C}
{Carroll} T.~A.,  {Kopf} M.,  {Strassmeier} K.~G.,   {Ilyin} I.,  2009, in
  {Strassmeier} K.~G.,  {Kosovichev} A.~G.,   {Beckman} J.~E.,  eds,  IAU
  Symposium Vol. 259, Cosmic Magnetic Fields: From Planets, to Stars and
  Galaxies. pp 633--644 (\mn@eprint {arXiv} {0903.1008}),
  \mn@doi{10.1017/S1743921309031469}

\bibitem[\protect\citeauthoryear{{Carroll}, {Strassmeier}, {Rice}  \&
  {K{\"u}nstler}}{{Carroll} et~al.}{2012}]{2012A&A...548A..95C}
{Carroll} T.~A.,  {Strassmeier} K.~G.,  {Rice} J.~B.,   {K{\"u}nstler} A.,
  2012, \mn@doi [\aap] {10.1051/0004-6361/201220215}, \href
  {http://adsabs.harvard.edu/abs/2012A%26A...548A..95C} {548, A95}

\bibitem[\protect\citeauthoryear{{Cernicharo} et~al.,}{{Cernicharo}
  et~al.}{1998}]{1998Sci...282..462C}
{Cernicharo} J.,  et~al., 1998, \mn@doi [Science]
  {10.1126/science.282.5388.462}, \href
  {http://adsabs.harvard.edu/abs/1998Sci...282..462C} {282, 462}

\bibitem[\protect\citeauthoryear{{D{\'{\i}}az}, {Gonz{\'a}lez}, {Levato}  \&
  {Grosso}}{{D{\'{\i}}az} et~al.}{2011}]{2011A&A...531A.143D}
{D{\'{\i}}az} C.~G.,  {Gonz{\'a}lez} J.~F.,  {Levato} H.,   {Grosso} M.,  2011,
  \mn@doi [\aap] {10.1051/0004-6361/201016386}, \href
  {http://adsabs.harvard.edu/abs/2011A%26A...531A.143D} {531, A143}

\bibitem[\protect\citeauthoryear{{Ekstr{\"o}m} et~al.,}{{Ekstr{\"o}m}
  et~al.}{2012}]{2012A&A...537A.146E}
{Ekstr{\"o}m} S.,  et~al., 2012, \mn@doi [\aap] {10.1051/0004-6361/201117751},
  \href {http://adsabs.harvard.edu/abs/2012A%26A...537A.146E} {537, A146}

\bibitem[\protect\citeauthoryear{{Ferrario}, {Pringle}, {Tout}  \&
  {Wickramasinghe}}{{Ferrario} et~al.}{2009}]{2009MNRAS.400L..71F}
{Ferrario} L.,  {Pringle} J.~E.,  {Tout} C.~A.,   {Wickramasinghe} D.~T.,
  2009, \mn@doi [\mnras] {10.1111/j.1745-3933.2009.00765.x}, \href
  {http://adsabs.harvard.edu/abs/2009MNRAS.400L..71F} {400, L71}

\bibitem[\protect\citeauthoryear{{Ferrario}, {Melatos}  \& {Zrake}}{{Ferrario}
  et~al.}{2015}]{2015SSRv..191...77F}
{Ferrario} L.,  {Melatos} A.,   {Zrake} J.,  2015, \mn@doi [\ssr]
  {10.1007/s11214-015-0138-y}, \href
  {http://adsabs.harvard.edu/abs/2015SSRv..191...77F} {191, 77}

\bibitem[\protect\citeauthoryear{{Gaia Collaboration,}, {Brown}, {Vallenari},
  {Prusti}, {de Bruijne}, {Mignard}  \& et al.}{{Gaia Collaboration,}
  et~al.}{2016}]{gaia}
{Gaia Collaboration,} {Brown} A. G.~A.,  {Vallenari} A.,  {Prusti} T.,  {de
  Bruijne} J. H.~J.,  {Mignard} F.,   et al. 2016, \aap, special Gaia volume,
  special volume, 1

\bibitem[\protect\citeauthoryear{{Giddings}}{{Giddings}}{1981}]{gid81}
{Giddings} J.~R.,  1981, PhD thesis, , University of London, (1981)

\bibitem[\protect\citeauthoryear{{Gonz{\'a}lez} \& {Levato}}{{Gonz{\'a}lez} \&
  {Levato}}{2006}]{2006A&A...448..283G}
{Gonz{\'a}lez} J.~F.,  {Levato} H.,  2006, \mn@doi [A\&A]
  {10.1051/0004-6361:20053177}, \href
  {http://adsabs.harvard.edu/abs/2006A%26A...448..283G} {448, 283}

\bibitem[\protect\citeauthoryear{{Gonz{\'a}lez} \& {Veramendi}}{{Gonz{\'a}lez}
  \& {Veramendi}}{2016}]{2016BAAA...58..105G}
{Gonz{\'a}lez} J.~F.,  {Veramendi} M.~E.,  2016, Boletin de la Asociacion
  Argentina de Astronomia La Plata Argentina, \href
  {http://adsabs.harvard.edu/abs/2016BAAA...58..105G} {58, 105}

\bibitem[\protect\citeauthoryear{{Grunhut} \& {Wade}}{{Grunhut} \&
  {Wade}}{2013}]{2013EAS....64...67G}
{Grunhut} J.~H.,  {Wade} G.~A.,  2013, in {Pavlovski} K.,  {Tkachenko} A.,
  {Torres} G.,  eds,  EAS Publications Series Vol. 64, EAS Publications Series.
  pp 67--74, \mn@doi{10.1051/eas/1364009}

\bibitem[\protect\citeauthoryear{{Hubrig}, {Ilyin}, {Sch{\"o}ller}  \& {Lo
  Curto}}{{Hubrig} et~al.}{2013}]{2013AN....334.1093H}
{Hubrig} S.,  {Ilyin} I.,  {Sch{\"o}ller} M.,   {Lo Curto} G.,  2013, \mn@doi
  [Astronomische Nachrichten] {10.1002/asna.201311948}, \href
  {http://adsabs.harvard.edu/abs/2013AN....334.1093H} {334, 1093}

\bibitem[\protect\citeauthoryear{{Hubrig} et~al.,}{{Hubrig}
  et~al.}{2014}]{2014A&A...564L..10H}
{Hubrig} S.,  et~al., 2014, \mn@doi [\aap] {10.1051/0004-6361/201423490}, \href
  {http://cdsads.u-strasbg.fr/abs/2014A%26A...564L..10H} {564, L10}

\bibitem[\protect\citeauthoryear{{Hurley}, {Tout}  \& {Pols}}{{Hurley}
  et~al.}{2002}]{2002MNRAS.329..897H}
{Hurley} J.~R.,  {Tout} C.~A.,   {Pols} O.~R.,  2002, \mn@doi [\mnras]
  {10.1046/j.1365-8711.2002.05038.x}, \href
  {http://adsabs.harvard.edu/abs/2002MNRAS.329..897H} {329, 897}

\bibitem[\protect\citeauthoryear{{Irrgang}, {Przybilla}, {Heber}, {B{\"o}ck},
  {Hanke}, {Nieva}  \& {Butler}}{{Irrgang} et~al.}{2014}]{2014A&A...565A..63I}
{Irrgang} A.,  {Przybilla} N.,  {Heber} U.,  {B{\"o}ck} M.,  {Hanke} M.,
  {Nieva} M.-F.,   {Butler} K.,  2014, \mn@doi [\aap]
  {10.1051/0004-6361/201323167}, \href
  {http://adsabs.harvard.edu/abs/2014A%26A...565A..63I} {565, A63}

\bibitem[\protect\citeauthoryear{{Kohoutek}, {Mayer}  \& {Lorenz}}{{Kohoutek}
  et~al.}{1999}]{1999A&AS..134..129K}
{Kohoutek} L.,  {Mayer} P.,   {Lorenz} R.,  1999, \mn@doi [\aaps]
  {10.1051/aas:1999128}, \href
  {http://cdsads.u-strasbg.fr/abs/1999A%26AS..134..129K} {134, 129}

\bibitem[\protect\citeauthoryear{{Krti{\v c}ka}, {Mikul{\'a}{\v s}ek}, {Zverko}
   \& {{\v Z}i{\v z}{\'n}ovsk{\'y}}}{{Krti{\v c}ka}
  et~al.}{2007}]{2007A&A...470.1089K}
{Krti{\v c}ka} J.,  {Mikul{\'a}{\v s}ek} Z.,  {Zverko} J.,   {{\v Z}i{\v
  z}{\'n}ovsk{\'y}} J.,  2007, \mn@doi [\aap] {10.1051/0004-6361:20066627},
  \href {http://adsabs.harvard.edu/abs/2007A%26A...470.1089K} {470, 1089}

\bibitem[\protect\citeauthoryear{{Kupka}, {Ryabchikova}, {Piskunov}, {Stempels}
   \& {Weiss}}{{Kupka} et~al.}{2000}]{kupka2000}
{Kupka} F.~G.,  {Ryabchikova} T.~A.,  {Piskunov} N.~E.,  {Stempels} H.~C.,
  {Weiss} W.~W.,  2000, Baltic Astronomy, \href
  {http://adsabs.harvard.edu/abs/2000BaltA...9..590K} {9, 590}

\bibitem[\protect\citeauthoryear{{Kurucz}}{{Kurucz}}{1993}]{1993KurCD..13.....K}
{Kurucz} R.,  1993, ATLAS9 Stellar Atmosphere Programs and 2 km/s grid.~Kurucz
  CD-ROM No.~13.~ Cambridge, Mass.: Smithsonian Astrophysical Observatory,
  1993., \href {http://adsabs.harvard.edu/abs/1993KurCD..13.....K} {13}

\bibitem[\protect\citeauthoryear{{Lanz} \& {Hubeny}}{{Lanz} \&
  {Hubeny}}{2007}]{2007ApJS..169...83L}
{Lanz} T.,  {Hubeny} I.,  2007, \mn@doi [ApJS] {10.1086/511270}, \href
  {http://adsabs.harvard.edu/abs/2007ApJS..169...83L} {169, 83}

\bibitem[\protect\citeauthoryear{{Lynds}, {Canzian}  \& {Oneil}}{{Lynds}
  et~al.}{1985}]{1985ApJ...288..164L}
{Lynds} B.~T.,  {Canzian} B.~J.,   {Oneil} Jr. E.~J.,  1985, \mn@doi [\apj]
  {10.1086/162775}, \href {http://adsabs.harvard.edu/abs/1985ApJ...288..164L}
  {288, 164}

\bibitem[\protect\citeauthoryear{{Morel} \& {Butler}}{{Morel} \&
  {Butler}}{2008}]{2008AA...487..307M}
{Morel} T.,  {Butler} K.,  2008, \mn@doi [\aap] {10.1051/0004-6361:200809924},
  \href {http://adsabs.harvard.edu/abs/2008A%26A...487..307M} {487, 307}

\bibitem[\protect\citeauthoryear{{Morel}, {Butler}, {Aerts}, {Neiner}  \&
  {Briquet}}{{Morel} et~al.}{2006}]{2006AA...457..651M}
{Morel} T.,  {Butler} K.,  {Aerts} C.,  {Neiner} C.,   {Briquet} M.,  2006,
  \mn@doi [\aap] {10.1051/0004-6361:20065171}, \href
  {http://adsabs.harvard.edu/abs/2006A%26A...457..651M} {457, 651}

\bibitem[\protect\citeauthoryear{{Morel} et~al.,}{{Morel}
  et~al.}{2014}]{2014Msngr.157...27M}
{Morel} T.,  et~al., 2014, The Messenger, \href
  {http://adsabs.harvard.edu/abs/2014Msngr.157...27M} {157, 27}

\bibitem[\protect\citeauthoryear{{Morel} et~al.,}{{Morel}
  et~al.}{2015}]{2015IAUS..307..342M}
{Morel} T.,  et~al., 2015, in {Meynet} G.,  {Georgy} C.,  {Groh} J.,   {Stee}
  P.,  eds,  IAU Symposium Vol. 307, New Windows on Massive Stars. pp 342--347
  (\mn@eprint {arXiv} {1408.2100}), \mn@doi{10.1017/S1743921314007054}

\bibitem[\protect\citeauthoryear{{Moss}}{{Moss}}{2001}]{2001ASPC..248..305M}
{Moss} D.,  2001, in {Mathys} G.,  {Solanki} S.~K.,   {Wickramasinghe} D.~T.,
  eds,  Astronomical Society of the Pacific Conference Series Vol. 248,
  Magnetic Fields Across the Hertzsprung-Russell Diagram. p.~305

\bibitem[\protect\citeauthoryear{{Nieva} \& {Przybilla}}{{Nieva} \&
  {Przybilla}}{2006}]{2006ApJ...639L..39N}
{Nieva} M.~F.,  {Przybilla} N.,  2006, \mn@doi [ApJL] {10.1086/501124}, \href
  {http://adsabs.harvard.edu/abs/2006ApJ...639L..39N} {639, L39}

\bibitem[\protect\citeauthoryear{{Nieva} \& {Przybilla}}{{Nieva} \&
  {Przybilla}}{2007}]{2007A&A...467..295N}
{Nieva} M.~F.,  {Przybilla} N.,  2007, \mn@doi [\aap]
  {10.1051/0004-6361:20065757}, \href
  {http://adsabs.harvard.edu/abs/2007A%26A...467..295N} {467, 295}

\bibitem[\protect\citeauthoryear{{Nieva} \& {Przybilla}}{{Nieva} \&
  {Przybilla}}{2008}]{2008AA...481..199N}
{Nieva} M.~F.,  {Przybilla} N.,  2008, \mn@doi [\aap]
  {10.1051/0004-6361:20078203}, \href
  {http://adsabs.harvard.edu/abs/2008A%26A...481..199N} {481, 199}

\bibitem[\protect\citeauthoryear{{Nieva} \& {Przybilla}}{{Nieva} \&
  {Przybilla}}{2012}]{2012A&A...539A.143N}
{Nieva} M.-F.,  {Przybilla} N.,  2012, \mn@doi [\aap]
  {10.1051/0004-6361/201118158}, \href
  {http://adsabs.harvard.edu/abs/2012A%26A...539A.143N} {539, A143}

\bibitem[\protect\citeauthoryear{{Nieva} \& {Sim{\'o}n-D{\'{\i}}az}}{{Nieva} \&
  {Sim{\'o}n-D{\'{\i}}az}}{2011}]{2011A&A...532A...2N}
{Nieva} M.-F.,  {Sim{\'o}n-D{\'{\i}}az} S.,  2011, \mn@doi [\aap]
  {10.1051/0004-6361/201116478}, \href
  {http://adsabs.harvard.edu/abs/2011A%26A...532A...2N} {532, A2}

\bibitem[\protect\citeauthoryear{{Przybilla}}{{Przybilla}}{2005}]{2005AA...443..293P}
{Przybilla} N.,  2005, \mn@doi [\aap] {10.1051/0004-6361:20053412}, \href
  {http://adsabs.harvard.edu/abs/2005A%26A...443..293P} {443, 293}

\bibitem[\protect\citeauthoryear{{Przybilla} \& {Butler}}{{Przybilla} \&
  {Butler}}{2001}]{2001AA...379..955P}
{Przybilla} N.,  {Butler} K.,  2001, \mn@doi [\aap]
  {10.1051/0004-6361:20011393}, \href
  {http://adsabs.harvard.edu/abs/2001A%26A...379..955P} {379, 955}

\bibitem[\protect\citeauthoryear{{Przybilla} \& {Butler}}{{Przybilla} \&
  {Butler}}{2004}]{2004ApJ...609.1181P}
{Przybilla} N.,  {Butler} K.,  2004, \mn@doi [\apj] {10.1086/421316}, \href
  {http://adsabs.harvard.edu/abs/2004ApJ...609.1181P} {609, 1181}

\bibitem[\protect\citeauthoryear{{Przybilla}, {Butler}, {Becker}, {Kudritzki}
  \& {Venn}}{{Przybilla} et~al.}{2000}]{2000AA...359.1085P}
{Przybilla} N.,  {Butler} K.,  {Becker} S.~R.,  {Kudritzki} R.~P.,   {Venn}
  K.~A.,  2000, \aap, \href
  {http://adsabs.harvard.edu/abs/2000A%26A...359.1085P} {359, 1085}

\bibitem[\protect\citeauthoryear{{Przybilla}, {Butler}, {Becker}  \&
  {Kudritzki}}{{Przybilla} et~al.}{2001}]{2001AA...369.1009P}
{Przybilla} N.,  {Butler} K.,  {Becker} S.~R.,   {Kudritzki} R.~P.,  2001,
  \mn@doi [\aap] {10.1051/0004-6361:20010164}, \href
  {http://adsabs.harvard.edu/abs/2001A%26A...369.1009P} {369, 1009}

\bibitem[\protect\citeauthoryear{{Przybilla}, {Nieva}  \& {Butler}}{{Przybilla}
  et~al.}{2011}]{2011JPhCS.328a2015P}
{Przybilla} N.,  {Nieva} M.-F.,   {Butler} K.,  2011, \mn@doi [Journal of
  Physics Conference Series] {10.1088/1742-6596/328/1/012015}, \href
  {http://adsabs.harvard.edu/abs/2011JPhCS.328a2015P} {328, 012015}

\bibitem[\protect\citeauthoryear{{Przybilla}, {Nieva}, {Irrgang}  \&
  {Butler}}{{Przybilla} et~al.}{2013}]{2013EAS....63...13P}
{Przybilla} N.,  {Nieva} M.~F.,  {Irrgang} A.,   {Butler} K.,  2013, in
  {Alecian} G.,  {Lebreton} Y.,  {Richard} O.,   {Vauclair} G.,  eds,  EAS
  Publications Series Vol. 63, EAS Publications Series. pp 13--23,
  \mn@doi{10.1051/eas/1363002}

\bibitem[\protect\citeauthoryear{{Przybilla} et~al.,}{{Przybilla}
  et~al.}{2016}]{2016A&A...587A...7P}
{Przybilla} N.,  et~al., 2016, \mn@doi [\aap] {10.1051/0004-6361/201527646},
  \href {http://adsabs.harvard.edu/abs/2016A%26A...587A...7P} {587, A7}

\bibitem[\protect\citeauthoryear{{Sota}, {Ma{\'{\i}}z Apell{\'a}niz},
  {Morrell}, {Barb{\'a}}, {Walborn}, {Gamen}, {Arias}  \& {Alfaro}}{{Sota}
  et~al.}{2014}]{2014ApJS..211...10S}
{Sota} A.,  {Ma{\'{\i}}z Apell{\'a}niz} J.,  {Morrell} N.~I.,  {Barb{\'a}}
  R.~H.,  {Walborn} N.~R.,  {Gamen} R.~C.,  {Arias} J.~I.,   {Alfaro} E.~J.,
  2014, \mn@doi [ApJs] {10.1088/0067-0049/211/1/10}, \href
  {http://cdsads.u-strasbg.fr/abs/2014ApJS..211...10S} {211, 10}

\bibitem[\protect\citeauthoryear{{Stasi{\'n}ska}}{{Stasi{\'n}ska}}{2009}]{2009elu..book....1S}
{Stasi{\'n}ska} G.,  2009, {What can emission lines tell us?}.
p.~1

\bibitem[\protect\citeauthoryear{{Torii} et~al.,}{{Torii}
  et~al.}{2011}]{2011ApJ...738...46T}
{Torii} K.,  et~al., 2011, \mn@doi [\apj] {10.1088/0004-637X/738/1/46}, \href
  {http://adsabs.harvard.edu/abs/2011ApJ...738...46T} {738, 46}

\bibitem[\protect\citeauthoryear{{Vrancken}, {Butler}  \& {Becker}}{{Vrancken}
  et~al.}{1996}]{1996AA...311..661V}
{Vrancken} M.,  {Butler} K.,   {Becker} S.~R.,  1996, \aap, \href
  {http://adsabs.harvard.edu/abs/1996A%26A...311..661V} {311, 661}

\bibitem[\protect\citeauthoryear{{Wade} et~al.,}{{Wade}
  et~al.}{2016}]{2016arXiv161002585W}
{Wade} G.~A.,  et~al., 2016, preprint, \href
  {http://adsabs.harvard.edu/abs/2016arXiv161002585W} {} (\mn@eprint {arXiv}
  {1610.02585})

\bibitem[\protect\citeauthoryear{{Yusef-Zadeh}, {Biretta}  \&
  {Geballe}}{{Yusef-Zadeh} et~al.}{2005}]{2005AJ....130.1171Y}
{Yusef-Zadeh} F.,  {Biretta} J.,   {Geballe} T.~R.,  2005, \mn@doi [\aj]
  {10.1086/432095}, \href {http://cdsads.u-strasbg.fr/abs/2005AJ....130.1171Y}
  {130, 1171}

\makeatother
\end{thebibliography}

\label{lastpage}

\end{document}